\documentclass[12pt, a4paper]{article}

\usepackage{amsmath}
\usepackage{amssymb}
\usepackage{tikz}
\usepackage{pgfplots}
\usepackage{slashed}
\usepackage{amsthm}
\usepackage{mathrsfs}
\usepackage{enumerate}
\usepackage{float}

\usepgfplotslibrary{fillbetween}

\theoremstyle{definition}

\newtheorem{thm}{Theorem}[section]
\newtheorem{prop}{Proposition}[section]
\newtheorem{rmk}{Remark}[section]
\newtheorem{lemma}{Lemma}[section]

\newcommand{\rmd}{\mathrm{d}}
\newcommand{\p}{\partial}
\newcommand{\too}{\longrightarrow}
\newcommand{\td}{\widetilde{\nabla}}

\newcommand{\Isom}{\text{Isom}}

\title{The Klein-Gordon equation on the hyperboloidal anti-de Sitter Schwarzschild black hole}
\author{Owain Salter Fitz-Gibbon}

\begin{document}

\maketitle
\begin{abstract}
	In this paper we establish energy decay for solutions to the Klein-Gordon equation on the positive mass hyperboloidal anti-de Sitter Schwarzschild black hole, subject to Dirichlet, Neumann and Robin boundary conditions at infinity, for a range of the (negative) mass squared parameter. To do so we use vector field methods with a renormalised energy to avoid divergences that would otherwise appear in the energy integrals. For another region of the parameter space, we use the existence of negative energy solutions to demonstrate linear instability.  
\end{abstract}
\section{Introduction}
\subsection{Background and motivation}
In recent years, the study of asymptotically anti de Sitter space-times has been brought into fashion, particularly in the physics community, because of the so-called AdS/CFT correspondence \cite{AdS/CFT}. These space-times have also attracted the attention of the mathematical community because they are believed to have interesting instability properties. Specifically, it is conjectured that arbitrarily small perturbations of AdS initial data can form black holes under the evolution of Einstein's equations: see for example \cite{AdSIntsability}, \cite{UniquenessGlobalDynamics}, \cite{WeaklyTurbulentInstability}, \cite{GravTurbulentInstability}. A special case of this conjecture for the Einstein-null dust system was proved by Moschidis in \cite{InstabilityEVAdS}, and for the Einstein-massless Vlasov system in \cite{InstabilityMaslessVlasov}.
\paragraph{}A natural first step when considering the full non-linear stability of solutions to the Einstein equations is to study the linear stability problem. That is, the boundedness and decay (or growth) of solutions to the linear Klein-Gordon equation,
\begin{equation}
\Box_g\psi +\frac{\alpha}{l^2}\psi = 0, 
\end{equation}
where $g$ is the metric of an asymptotically AdS space-time with AdS radius $l$. Notice that as written above, and with the $(-+++)$ convention used here for the metric, a positive $\alpha$ corresponds to a negative ``mass squared parameter''. \paragraph{}Since asymptotically AdS spaces are not globally hyperbolic, the well-posedness of this equation requires that boundary conditions be imposed at null infinity. This amounts to imposing restrictions on the asymptotic behaviour of the solution as $r\too\infty$. When Breitenlohner and Freedman solved the equation on the exact AdS space-time \cite{GESG}, they found that there were two branches of the solution, decaying at different rates towards infinity, leading to three natural types of boundary conditions that may be imposed, referred to as Dirichlet, Neumann, and Robin conditions. Dirichlet conditions require that the more slowly decaying branch of the solution vanishes, and Neumann that the more quickly decaying branch vanishes. Robin conditions require that some combination of the two branches vanish. The well-posedness of this equation has been established for all three types of boundary conditions, with different ranges of $\alpha$. The work of Holzegel \cite{WellPosednessGH} and Vasy \cite{MicrolocalNonsense} established well-posedness in the range $\alpha<9/4$ with Dirichlet boundary conditions (the latter using techniques of microlocal analysis), and later Warnick \cite{MassiveWaveEqn} proved well-posedness in the range $0<\alpha<9/4$ for Dirichlet boundary conditions and $5/4<\alpha<9/4$ for Robin and Neumann boundary conditions. Crucially, the Neumann and Robin cases were dealt with using the notion of the twisted derivative, which allows certain energy integrals to be renormalised, and which will play an important role in the present paper.
\paragraph{}An interesting feature of the asymptotically AdS analogue of the Schwarzschild black hole is that the horizon geometry need not be spherical; it can also be a flat plane (or the flat torus $\mathbb{R}^2/\mathbb{Z}^2$) or a hyperbolic plane (or any genus $g\geq2$ surface which can be obtained as a quotient of the hyperbolic plane by a freely acting discrete group of isometries). In the physics literature, black holes with a `non-trivial' topology are referred to as \emph{topological black holes} (see for example \cite{TopBHs}). In \cite{TopBHPP}, Mann  showed that quantum mechanical pair production of such black holes is possible, and in \cite{TopBHGravCollapse} and \cite{GravCollapseLemos} Mann and Smith, and Lemos respectively proved that such black holes can be formed classically by the gravitational collapse of a dust cloud. It is believed that topological black holes are of interest in understanding theories of quantum gravity which include topology changing processes \cite{TopBHs}.

\paragraph{}The linear Klein-Gordon equation on the spherical and toroidal (or planar) black holes has already been studied in a number of works. In \cite{HolSmul}, Holzegel and Smulevici proved that for solutions to the Klein-Gordon equation on Kerr-AdS, a non-degenerate energy decays slowly in time (as an inverse power of the logarithm). In \cite{Boundedness&Growth}, Holzegel and Warnick studied the linear stability (in the sense of uniform boundedness of solutions of the Klein-Gordon equation) of stationary AdS black holes in general, and of the spherically symmetric AdS Schwarzschild black hole in particular. The full non-linear (orbital and asymptotic) stability of the spherical AdS Schwarzschild black hole was proved in \cite{SphericalStability} within the class of spherical symmetry. In the toroidal case, energy decay for solutions to the linear equation (this time at a polynomial rate) was proved in \cite{ToroidalAdS} by Dunn and Warnick. It was also proved here that this decay rate can only hold with a loss of derivative, due to the existence of null geodesics which remain outside the horizon for arbitrarily long times. The non-linear stability (again in the class of toroidal symmetry) was proved by the same authors in \cite{ToroidalStability}. 
\paragraph{}Another interesting feature of the AdS Schwarzschild black hole is that in the hyperboloidal case, the black hole persists for a range of non-positive mass parameter, and that there exists a mass $M^{\text{ext}}<0$ for which the black hole is extremal. This fact makes the hyperboloidal black hole a useful toy model when studying extremal or near-extremal Reissner-Nordstr{\"o}m-AdS and Kerr-AdS black holes, as in \cite{SFCondensation}. We also note that when $M=0$ and $\alpha=2$ (the conformally coupled case), there exists a non-zero, time-independent solution of the equation. This suggests that it might be possible to prove the existence of a hairy black hole solution of the coupled Einstein-Scalar field system.  
\paragraph{}In $4+1$ dimensions the linear and non-linear stability of the analogue of this space-time was investigated numerically by Dias, Monteiro, Reall, and Santos in \cite{SFCondensation}, who used this as a toy model to study the instability of rotating black holes to scalar field condensation. They found numerically that the black hole is unstable to the condensation of a scalar field, for various ranges of the mass squared parameter $\alpha$ and black hole mass $M$. The linear stability in arbitrary $d$ dimensions has been studied in the $M=0$ case by Belin and Maloney in \cite{InstabilityTopBH}, and applied to the stability of Conformal Field Theories on negatively curved compact spaces. In particular, in $3+1$ dimensions they found growing mode solutions in the case $5/4<\alpha<2$, but only when Neumann boundary conditions are imposed; these results are in agreement with the arguments in this paper. No growing modes were found for Dirichlet boundary conditions, leaving open the question of whether a decay result can be proved for the $M=0$ case when Dirichlet boundary conditions are imposed. 
\paragraph{}From the point of view of the AdS/CFT correspondence, the instability of AdS black holes to condensation by (charged) scalar fields can be dual to superconducting phase transitions in a field theory on the boundary \cite{HolographicSuperconductors}. The instability of the hyperboloidal AdS black hole to condensation by an uncharged scalar field (first shown numerically in \cite{SFCondensation}) is linked to the non-analyticity of R\'{e}nyi entropy of a CFT in \emph{flat} space \cite{RenyiEntropy}. Specifically, the instability of the black hole to scalar hair implies that the R\'{e}nyi entropy $S_n$ is non-analytic as a function of $n$.
\subsection{Contents of the paper}
The subject of this paper is the Klein-Gordon equation on the $3+1$ dimensional hyperboloidal black hole. Throughout, we work in coordinates $\left(t^*,r,\sigma,\phi\right)$ which are regular at the horizon (and where $\left(\sigma,\phi\right)$ are coordinates on the hyperbolic plane).
\paragraph{}We begin by defining the space-time we will study in section 2. Section 3 examines some properties of null geodesics in this space-time. Section 4 discusses the Klein-Gordon equation, recalling results about well-posedness and proving energy boundedness statements. Our two main results on linear instability and energy decay are contained in sections 5 and 6 respectively.
\subsubsection{Energy Boundedness}
The first important result in this paper occurs in section 4, and concerns the boundedness of the \emph{degenerate} energy $E_{t^*}[\psi]$, defined by equation \eqref{RenormalisedEnergy}. (Here by ``energy'', we mean a suitably chosen quadratic form of the field and its first derivatives). This result says that when the black hole mass $M\geq0$, $E_{t^*}[\psi]\leq E_{0}[\psi]$, but if $M\leq0$, then $E_{t^*}[\psi]\geq E_{0}[\psi]$. In other words, if the black hole mass is \emph{non-negative}, then the energy is bounded \emph{above} by its initial value, whereas if the black hole mass is \emph{negative}, then it is bounded \emph{below} by its initial value. In particular, a solution with \emph{negative} (resp. \emph{positive}) initial energy when $M\geq0$ (resp. $\leq0$) has energy which remains bounded away from zero. In section 5 it is explained how the existence of a solution with energy bounded away from zero implies linear instability, and such solutions are constructed in certain regions of parameter space.
\subsubsection{Regions of stability and instability}\label{Summary}
Arguments from \cite{Boundedness&Growth} and \cite{Quasinormalmodes} show that if we can find a solution of the equation, the energy of which remains bounded away from zero, then we must have linear instability (for a precise definition of what is meant by linear instability in this case, see section \ref{NegEnergyInst}). The region in the parameter space for which this is possible is illustrated here. To most conveniently describe the different regions, we parameterise the black hole by the horizon radius $r_+=r_+(M)$ and the Klein-Gordon mass by $\kappa=\sqrt{9/4-\alpha}$. Note that the range $2\leq\alpha<9/4$ corresponds to $1/2\geq\kappa>0$. 

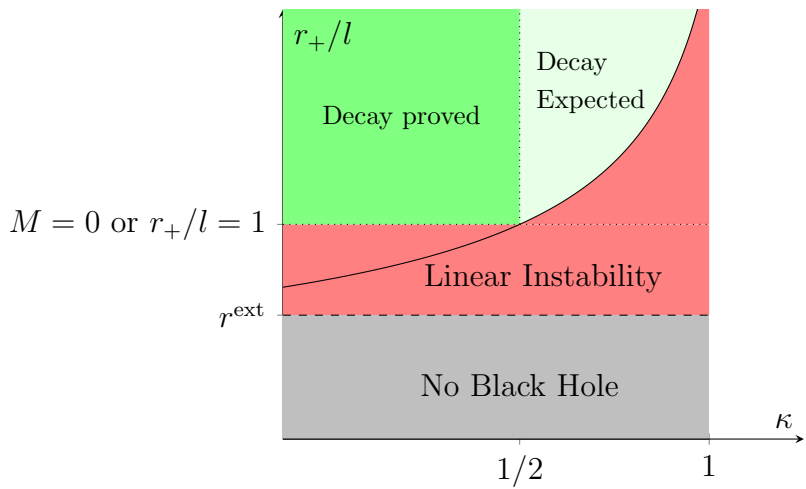
\begin{figure}[H]
	\centering
	\begin{tikzpicture}
	\begin{axis}[axis lines = center,
	xlabel = $\kappa$,
	ylabel = {$r_+/l$},
	xtick={0.5,0.9},
	ytick={0,0.577,1},
	xticklabels={$1/2$,$1$},
	yticklabels={$0$,$r^{\text{ext}}$,$M=0\text{ or }r_+/l=1$},
	xmin=0,
	xmax=1.1,
	ymin=0,
	ymax=2,
	]	
	\addplot[name path=ThresholdOne, black, smooth, domain = 0:0.5,
	]
	{1/sqrt(2-2*x)};
	\addplot[name path=ThresholdTwo, black, smooth, domain = 0.5:0.9,
	]
	{1/sqrt(2-2*x)};
	\addplot[name path=Extremality, black, dashed, domain = 0:0.9,
	]
	{0.577};
	\addplot[name path=Top, black, dashed, domain = 0:0.9,
	]
	{2.1};
	\addplot[name path=Bottom, black, domain = 0:0.9,
	]
	{0};
	\addplot[name path=MequalszeroOne, black, dotted, domain = 0:0.5,
	]
	{1};
	\addplot[name path=MequalszeroTwo, black, dotted, domain = 0.5:0.9,
	]
	{1};
	\addplot[name path=kappaequalhalf,dotted, black]
	coordinates{(0.5,1) (0.5,2)};
	\addplot[red!50] fill between[of=MequalszeroTwo and ThresholdTwo];
	\addplot[red!50] fill between[of=Extremality and MequalszeroOne];
	\addplot[red!50] fill between[of=Extremality and MequalszeroTwo];
	
	\addplot[green!50] fill between[of=MequalszeroOne and Top];
	\addplot[gray!50] fill between[of=Bottom and Extremality];
	\addplot[green!10] fill between[of=ThresholdTwo and kappaequalhalf];
	\node at (axis cs:0.5,0.25) {No Black Hole};
	\node(-align)[align=right] at (axis cs:0.55,0.75) {Linear Instability};
	\node at (axis cs:0.25,1.5) {\footnotesize Decay proved};
	\node at (axis cs:0.2, 0.9){} ;
	\node(-align)[align=left] at (axis cs:0.65, 1.66) {\footnotesize Decay\\ \footnotesize Expected};
	\end{axis}
	\end{tikzpicture}
	\caption{The regions of linear stability and instability for Neumann boundary conditions}
	\label{fig:PrettyPicture}
\end{figure}

\paragraph{}A summary of our results, at least for Neumann boundary conditions, is expressed in the following diagram of the $\left(\kappa, r_+\right)$ plane (Figure \ref{fig:PrettyPicture}). Note that $r_+\geq l$ corresponds to $M\geq0$. The region of linear instability is given by the inequality 
\begin{equation}\label{InstabilityThresholdr_+}
\frac{r_+^2}{l^2}<1-\frac{1}{2}\cdot\frac{1-2\kappa}{1-\kappa},
\end{equation}
when $r_+/l\geq1$, together with all points where $r_+/l\leq1$.
We would expect the threshold of instability for Dirichlet boundary conditions to be higher than for Neumann boundary conditions, but the precise region of stability for Dirichlet boundary conditions is yet to be determined. 
\subsubsection{Energy decay}
In the final section of the paper, we follow the methods of \cite{Boundedness&Growth} and particularly \cite{ToroidalAdS} to prove a polynomial decay rate for a non-degenerate energy of solutions of the Klein-Gordon equation when $M>0$ and $1/2\geq \kappa>0$. Note that this includes the conformally coupled case $\kappa=1/2$. 
\paragraph{}In what follows, suppose that $\psi$ is a solution to the Klein-Gordon equation with $1/2\geq\kappa>0$ on the exterior of the $M>0$ hyperboloidal AdS Schwarzschild black hole (in $3+1$ dimensions), obeying Dirichlet, Neumann or Robin boundary conditions (with the Robin function $\beta$ non-negative and independent of time) at infinity, and let $\mathcal{E}[\psi]$ be the \emph{non-degenerate} energy density defined by equation \eqref{RenormalisedEnergyDensity}. (When $M>0$ and $1/2\geq\kappa>0$, this energy is positive definite.) We begin by proving that this non-degenerate energy is bounded above by (a multiple of) its initial value.
\begin{thm}\label{IntroBoundedness}
	For any time $T>0$\footnote{Recall that if $f$, $g$ are non-negative functions, then the statement $f(x)\lesssim g(x)$ means that there exists a constant $C>0$ such that $f(x)\leq Cg(x)$ for all $x$.},
	\begin{equation}
		\int_{t^*=T}\mathcal{E}[\psi] \lesssim \int_{t^*=0}\mathcal{E}[\psi]. 
	\end{equation} 
\end{thm} 
The next step is to prove an integrated decay, or Morawetz, estimate. 
\begin{thm}
	 	For any time $T>0$,
	 \begin{equation}
	 \int_{0\leq t^*\leq T}\frac{\mathcal{E}[\psi]}{r^2} \lesssim \int_{t^*=0}\mathcal{E}[\psi]. 
	 \end{equation}
\end{thm}
\begin{rmk}
	The weaker weight on the left hand side of this inequality in fact need only be applied to the $t^*$ derivative. It is interesting to note that in this estimate the integrand on the left hand side is $1/r^2$ times the integrand on the right. This is better than in the toroidal case, which requires a factor of $1/r^3$ on the left. Moreover, in that case the less favourable weights apply to the derivatives tangent to the torus, as well as the time derivatives. See \cite{ToroidalAdS}.
\end{rmk}
This loss in weight on the left hand side can be removed, provided we control the initial energy of $\p_{t^*}\psi$.
\begin{thm}
		For any time $T>0$,
	\begin{equation}
	\int_{0\leq t^*\leq T}\mathcal{E}[\psi] \lesssim \int_{t^*=0}\left(\mathcal{E}[\psi]+\mathcal{E}[\p_{t^*}\psi]\right). 
	\end{equation}
\end{thm}
Finally, using a result from \cite{LFEAdS} (subsequently used in \cite{ToroidalAdS}) which involves combining the Morawetz estimate above with a quantitative version of the redshift effect taken from \cite{Quasinormalmodes}, we are able to conclude the following quantitative energy decay estimate.
\begin{thm}
 For any time $T>0$, and natural number $n\in \mathbb{N}$,
 \begin{equation}
 \int_{t^*=T}\mathcal{E}[\psi] \lesssim \frac{1}{\left(1+T\right)^n}\int_{t^*=0}\sum_{k=0}^{n}\mathcal{E}[\p_{t^*}^k\psi]. 
 \end{equation}
\end{thm}
\begin{rmk}
	In all of the above theorems, the implicit constant depends on the space-time parameters $(M,l)$ and the Klein-Gordon parameter $\alpha$ but is independent of $\psi$ and $T$.
\end{rmk}

\paragraph{}We can contrast the results above with numerical work done in \cite{SFCondensation} which looks at the same problem in $4+1$ dimensions and finds evidence of a linear instability occurring for various values of $\alpha$ and $M$, including a range of positive $M$. 

\subsection{Acknowledgements}
I would like to thank Claude Warnick for introducing me to this problem, and for many useful suggestions throughout. I am also grateful to Mihalis Dafermos, Jorge Santos, and Fred Alford for their helpful comments.
\section{The space-time}

The metric

\begin{equation}\label{Metric}
	g = - \left(k - \frac{2M}{r} + \frac{r^2}{l^2}\right)\rmd t^2 + \frac{\rmd r^2}{\left(k - \frac{2M}{r} + \frac{r^2}{l^2}\right)} + r^2 \rmd \Omega_k^2,
\end{equation}
 is a solution of the vacuum Einstein equations with a negative cosmological constant, referred to as the anti-de Sitter Schwarzschild black hole \cite{KottlerSoln}, \cite{AllegedPlanarAdSGuy}. Here $M$ is the black hole mass, $l$ is the AdS radius, $k=-1,0,1$ and $\rmd \Omega_k^2$ is the metric on a two-dimensional surface of constant sectional curvature $k$. In this paper, we will look at the Klein-Gordon equation associated to this metric,
\begin{equation}\label{KGeqn}
	\Box_g\psi +\frac{\alpha}{l^2}\psi = 0, 
\end{equation}
in the particular case that $k=-1$, on the exterior of the black hole, and where $\alpha$ satisfies the \emph{Breitenlohner-Friedman} bound $0<\alpha<9/4$ \cite{GESG}. The cases when $k=1,0$ have been studied already in \cite{Boundedness&Growth} and \cite{ToroidalAdS} respectively. 
\begin{rmk}
	 Unlike the cases $k=0,1$ when $k=-1$ there is a black hole for non-positive mass, provided that $M\geq M^{\text{ext}}:=-l/\left(3\sqrt{3}\right)$. When equality holds in the above, the black hole is extremal (see Section 3 in \cite{SFCondensation}). The decay results proved in this paper do not hold for non-positive $M$, however see sections \ref{Summary} and \ref{Instability} for results about linear instability in this case.
\end{rmk}

Putting $k=-1$ in \eqref{Metric} we see that there is a coordinate singularity at any value of $r$ where the polynomial  

\begin{equation}
p(r) = r^3 - l^2r -2Ml^2
\end{equation}
vanishes.
\begin{figure}[H]
	\centering
	\begin{tikzpicture}
	\begin{axis}[axis lines = center,
	xlabel = $r$,
	ylabel = {$p(r)$},
	ymin=-15,
	xtick={-1.633, 1.633,3},
	ytick=\empty,
	xticklabels={$-\frac{\sqrt{3}}{3}l$,$\frac{\sqrt{3}}{3}l$,}
	]	
	\addplot[smooth, domain = -2:4,
	]
	{x^3 - 8*x -3};
	\node[label={300:{$r_+$}}] at (axis cs:3,0){} ;
	\end{axis}
	\end{tikzpicture}
	\caption{The function $p(r)$, with $M>0$}
\end{figure}
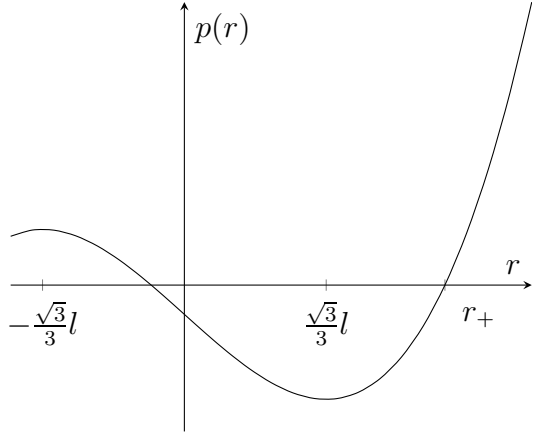
 
Note that $p'(r) = 3r^2 - l^2$, so the stationary points of $p$ are precisely the points $r= \pm \frac{\sqrt{3}}{3}l$. Since $p''(r) = 6r$ for all $r$, we see that $r=\frac{\sqrt{3}}{3}l$ is a local minimum for $p$ and $r=-\frac{\sqrt{3}}{3}l$ is a local maximum. Since $p\left(\frac{\sqrt{3}}{3}l\right)< 0$, it follows that $p$ has precisely one root, say $r_+,$ in the interval $\left(\frac{\sqrt{3}}{3}l, \infty \right)$. In fact, since $p(l)<0$, it must be the case that $r_+>l$. Since $p(0)<0$ and $p$ has no stationary points in the interval $\left(0,\frac{\sqrt{3}}{3}l \right)$, it follows that $r_+$ is the unique positive root of $p$.
\begin{rmk}
	If $M<0$, then $p$ need not have a unique positive root. In this case, we define $r_+$ to be the largest positive root. The extremal value of the mass $M^{\text{ext}}=-l/(3\sqrt{
	3})$ corresponds to a horizon radius $r_+^{\text{ext}}=l/\sqrt{3}$, and $M=0$ corresponds to $r=l$.
\end{rmk} 
\paragraph{}It's possible to parameterise the metric either by $(M,l)$ or by $(r_+,l)$. When switching between the two it is useful to have a picture of the relation between $M$ and $r_+$, given by
\begin{equation}
	M=\frac{r_+}{2}\left(\frac{r_+^2}{l^2}-1\right).
\end{equation} 

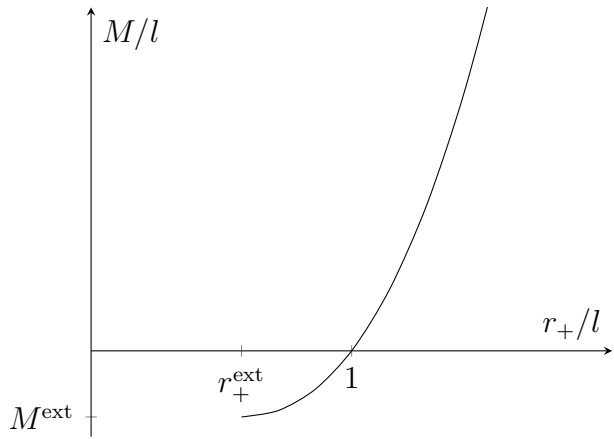
\begin{figure}[h]
	\centering
	\begin{tikzpicture}
	\begin{axis}[axis lines = center,
	xlabel = $r_+/l$,
	ylabel = {$M/l$},
	xtick={0.57735,1},
	ytick={-0.19245},
	xticklabels={$r_+^{\text{ext}}$,1},
	yticklabel={$M^{\text{ext}}$},
	xmin=0, 
	xmax=2,
	ymin=-0.25,
	ymax=1
	]	
	\addplot[smooth, domain = 0.57735:4,
	]
	{x^3/2 - x/2};
	\end{axis}
	\end{tikzpicture}
	\caption{The black hole mass $M$ as a function of the horizon radius $r_+$}
\end{figure}

As in the Schwarzschild metric, there is a coordinate singularity at precisely one value of the radius, $r=r_+$, representing the black hole horizon and, as in that case, we can perform a change of coordinates which allows the metric to be extended to part of the region $0<r<r_+$. Define a new coordinate $t^*$ by 

\begin{equation}
\rmd t^* = \rmd t + \frac{2M}{r} \frac{1}{\left(-1+\frac{r^2}{l^2}\right)} \frac{1}{\left(-1 -\frac{2M}{r}+\frac{r^2}{l^2}\right)}\,\rmd r.
\end{equation}
After a short calculation, we see that the metric in $(t^*,r)$ coordinates is 

\begin{equation}\label{MetricReg}
	g=-\left(-1-\frac{2M}{r}+\frac{r^2}{l^2}\right)(\rmd t^*)^2 + \frac{4M}{r}\frac{1}{\left(-1+\frac{r^2}{l^2}\right)}\rmd t^*\rmd r + \frac{-1+\frac{2M}{r}+\frac{r^2}{l^2}}{\left(-1+\frac{r^2}{l^2}\right)^2}\rmd r^2 + r^2\rmd \Omega_{-1}^2.
\end{equation}
For later convenience, we will define the two functions
\begin{equation}
f(r)=-1-\frac{2M}{r}+\frac{r^2}{l^2},
\end{equation} 
and 
\begin{equation}
g(r)=-1+\frac{2M}{r}+\frac{r^2}{l^2}.
\end{equation}
Since $r_+>l$, this metric is regular at the horizon $r=r_+$.
We will now formally define the Lorentzian manifold (with boundary) $(\mathcal{M},g)$ to be

\begin{equation}\label{Manifold}
\mathcal{M} = \mathbb{R}_{t^*} \times \mathbb{R}_{r\geq r_+} \times \mathbb{H}^2
\end{equation}
with the metric \eqref{MetricReg}, where $\mathbb{H}^2$ is the two dimensional hyperbolic plane, with metric

\begin{equation}
	\rmd \Omega_{-1}^2 = \rmd \sigma^2 +\sinh^2(\sigma)\rmd \phi^2
\end{equation} 
where $0<\sigma<\infty$ and $\phi$ is periodic with period $2\pi$.
The non-zero components of the inverse metric are 
\begin{equation}
\begin{split}
g^{t^*t^*} = -\frac{-1+\frac{2M}{r} + \frac{r^2}{l^2}}{\left(-1+\frac{r^2}{l^2}\right)^2},&\, g^{t^*r}=g^{rt^*}=\frac{2M}{r} \frac{1}{-1+\frac{r^2}{l^2}},\, g^{rr}=\left(-1-\frac{2M}{r}+\frac{r^2}{l^2}\right),
\\& g^{\sigma\sigma}=\frac{1}{r^2},\,g^{\phi\phi}=\frac{1}{r^2\sinh^2\left(\sigma\right)}.
\end{split}
\end{equation}
The volume form is 
\begin{equation}
\rmd \text{Vol} = \sqrt{-g}\,\rmd t^* \,\rmd r\,\rmd \sigma\,\rmd \phi=r^2\,\rmd t^* \,\rmd r\,\rmd \omega,
\end{equation}
where $\rmd \omega = \sinh^2 \sigma \,\rmd \sigma \,\rmd\phi$ is the volume form of the hyperbolic plane. It will be convenient to let $\slashed\nabla$ denote the covariant derivative on the hyperbolic plane (where we include the factor of $r^2$ in the metric).
\begin{rmk}
	In the definiton of the manifold \eqref{Manifold} above, we could replace the hyperbolic plane $\mathbb{H}^2$ by any quotient $\mathbb{H}^2/\Gamma$, where $\Gamma$ is a freely-acting discrete subgroup of $\Isom(\mathcal{M})$. The resulting manifold would have the same metric but a different topology. In particular, $\mathbb{H}^2/\Gamma$ may be chosen to be compact, whereas $\mathbb{H}^2$ is not. For simplicity, we will consider a compact quotient so that we don't need to worry about convergence of integrals over the hyperbolic planes. However we could consider the non-compact case by assuming sufficient decay in the hyperbolic directions.  
\end{rmk}
\subsection{Some hypersurfaces and their normals}
In what follows it will be convenient to define some hypersurfaces and write down their unit normal vectors. 
\begin{itemize}
	\item 
	Let $\Sigma_{t^*}$ denote the surface of constant $t^*$. The future-directed unit normal is 
	\begin{equation}
	n_a= -\frac{1}{\sqrt{-g^{t^*t^*}}} \, \left(\rmd t^* \right)_a,
	\end{equation}
	or with the index raised
	\begin{equation}\label{formulanup}
	n^a= \sqrt{-g^{t^*t^*}} \left( \frac{\p }{\p t^*}\right)^a- \frac{g^{rt}}{\sqrt{-g^{t^*t^*}}}\left(\frac{\p }{\p r}\right)^a.
	\end{equation}
	The volume element is 
	\begin{equation}
	\rmd S_{\Sigma_{t^*}} = \sqrt{-g^{t^*t^*}}r^2\, \rmd r\, \rmd \omega.
	\end{equation}
	\item 
	Let $\tilde{\Sigma}_{r}$ denote the surface of constant $r$. The unit normal is 
	\begin{equation}
	m_a=\frac{1}{\sqrt{g^{rr}}}\,\left(\rmd r\right)_a,
	\end{equation}
	or with the index raised 
	\begin{equation}\label{formulamup}
	m^a = \frac{g^{t^*r}}{\sqrt{g^{rr}}}\left(\frac{\p}{\p t^*}\right)^a + \sqrt{g^{rr}} \left(\frac{\p}{\p r}\right)^a.
	\end{equation}
	The volume element is
	\begin{equation}
	\rmd S_{\widetilde{\Sigma}_r} = \sqrt{g^{rr}} r^2 \, \rmd t^*\, \rmd \omega
	\end{equation}
	Notice that $g^{rr} \too 0$ as $r\too r_+$, and $g^{t^*r} \too \frac{2M}{r_+} \frac{1}{-1+\frac{r^2}{l^2}}$ as $r \too r_+$. It follows that $m^a$ becomes singular at the horizon $r=r_+$, but the product $m^a \rmd S_{\tilde{\Sigma}_r}$ is well behaved, and so defines a natural vector volume form on the horizon. 
\end{itemize} 

\section{Null geodesics}
Before beginning the study of the equation, we will look at the null geodesics of the space-time $(\mathcal{M},g)$. Note that a detailed study of the geodesics of the spherical AdS Schwarzschild black hole has been made in \cite{GeodesicsSchAdS}.

\subsection{The equation of motion}
In this section, we will show that null geodesics in $\left(\mathcal{M},g\right)$ obey a one dimensional potential equation. As usual, we begin by using the symmetries of the Hamiltonian to find conserved quantities. Let $\gamma$ be an affinely parameterised null geodesic (say with affine parameter $s$), and denote the coordinates of $\gamma$ by $\left(x^\mu\right)$. The geodesic Lagrangian is
\begin{equation}
L=g_{\mu\nu}\dot{x}^\mu\dot{x}^\nu.
\end{equation}
The momentum conjugate to the position variable $x^\mu$ is 
\begin{equation}\label{momenta}
p_\mu=\frac{\p L}{\p \dot{x}^\mu}=2g_{\mu\nu}\dot{x}^\nu.
\end{equation}
That is,
\begin{equation}
\begin{split}
	p_{t^*}&=-2f(r)\dot{t}^*+\frac{4M}{r}\frac{1}{\left(-1+\frac{r^2}{l^2}\right)}\dot{r},
	\\p_r&=\frac{4M}{r}\frac{1}{\left(-1+\frac{r^2}{l^2}\right)}\dot{t}^*+\frac{2g(r)}{\left(-1+\frac{r^2}{l^2}\right)^2}\dot{r},
	\\p_\sigma&=2r^2\dot{\sigma}, \text{ and}
	\\p_\phi&=2r^2\sinh^2\left(\sigma\right)\dot{\phi}.
\end{split}
\end{equation}
Since $g$ is independent of $t^*$, the conjugate momentum
$p_{t^*}$ is conserved along geodesics. Define
\begin{equation}\label{eq:defE}
E = \frac{p_{t^*}}{2}=-f(r)\dot{t}^* + \frac{2M}{r} \frac{1}{-1 + \frac{r^2}{l^2}}\dot{r}=\text{const}.
\end{equation}
Inverting the relation \eqref{momenta}, we get
\begin{equation}
	\dot{x}^\mu=\frac{1}{2}g^{\mu\nu}p_\nu.
\end{equation}
The Hamiltonian is 
\begin{equation}\label{Hamiltonian}
H=p_\mu\dot{x}^\mu-L=L.
\end{equation}
Plugging \eqref{momenta} in to \eqref{Hamiltonian}, we obtain the formula for the Hamiltonian in terms of the conjugate momenta,
\begin{equation}
\begin{split}
	H&=\frac{1}{4}g^{\mu\nu}p_\mu p_\nu,
	\\&=\frac{1}{4}\left(-\frac{g(r)}{\left(-1+\frac{r^2}{l^2}\right)^2}p_{t^*}^2+f(r)p_r^2+\frac{1}{r^2}\left(p_\sigma^2+\frac{1}{\sinh^2(\sigma)}p_\phi^2\right)\right).
\end{split}
\end{equation}
It is then easy to see that
\begin{equation}
	\left\{p_\sigma^2+\frac{1}{\sinh^2(\sigma)}p_\phi^2,H\right\}=0,
\end{equation}
and so 
\begin{equation}
	h^2=\frac{1}{4}\left(p_\sigma^2+\frac{1}{\sinh^2(\sigma)}p_\phi^2\right)
\end{equation}
is conserved along geodesics. In terms of the velocities, rather than the momenta, we see that
\begin{equation}
	h^2=r^4\left(\dot{\sigma}^2+\sinh^2(\sigma)\dot{\phi}^2\right).
\end{equation}
Finally, $H$ itself is conserved, and since $\gamma$ is null, is in fact equal to zero. Putting this all together, we get
\begin{equation}
	0=H=\frac{E^2-\dot{r}^2}{f(r)}+\frac{h^2}{r^2},
\end{equation}
or after tidying up slightly,
\begin{equation}\label{EOM}
\dot{r}^2 +\frac{h^2}{r^2}f(r) = E^2.
\end{equation}
\begin{rmk}
	Notice that the motion in the hyperbolic plane is decoupled from the radial motion, except for the presence of the constant $h^2$ in the equation of motion \eqref{EOM}.
\end{rmk}
 
\subsection{Gravitational attraction of the black hole}
If we differentiate \eqref{EOM} with respect to the affine parameter $s$ we get 
\begin{equation}
\dot{r}\left(2\ddot{r} + \frac{\rmd}{\rmd r}\frac{h^2f(r)}{r^2}\right)=0.
\end{equation}
It is easy to check that $f(r)/r^2$ is a strictly increasing function of $r$, which tends to $1/l^2$ as $r \too \infty$. Therefore, provided that $\dot{r} \neq 0$, it follows that $\ddot{r}<0$. That is, a light ray accelerates radially inwards towards the black hole, and the acceleration decreases to zero as $1/r^3$ as $r$ increases to infinity.
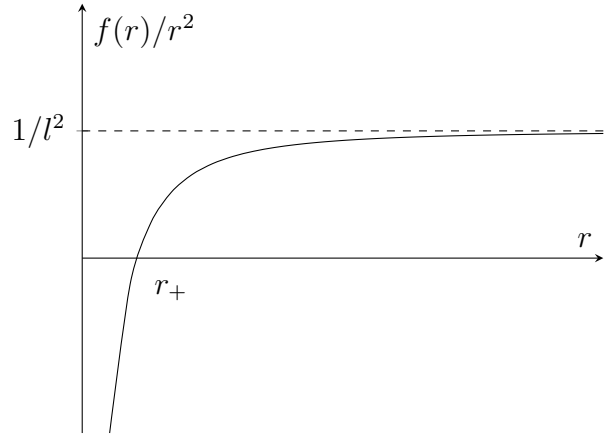
\begin{figure}[h]
	\centering
	\begin{tikzpicture}\label{f/r^2}
	\begin{axis}[axis lines = center,
	xlabel = $r$,
	ylabel = {$f(r)/r^2$},
	xtick=\empty,
	ytick={100},
	yticklabels={$1/l^2$},
	ymax=200,
	]	
	\addplot[smooth, domain = 2:20,
	]
	{100*(1 - 8/x^2 -3/x^3)};
	\addplot[dashed, smooth, domain = 1:20,
	]
	{100};
	\node[label={300:{$r_+$}}] at (axis cs:3,0){} ;
	\end{axis}
	\end{tikzpicture}
	\caption{The function $f(r)/r^2$}
\end{figure} 
\subsection{Absence of constant $r$ null geodesics}
As mentioned above, $f(r)/r^2$ is a strictly increasing function of $r$, and $f(r)/r^2 \too 1/l^2$ as $r \too \infty$. Since $f(r_+)=0$, it follows that whenever $h^2/l^2 > E^2$ there is a unique solution $r_0>r_+$ to the equation $h^2f(r)/r^2=E^2$. We might then expect there to be a null geodesic with $r(s) = r_0$ (and $\dot{r}(s)=0$) for all $s$. In fact this cannot happen, and $\dot{r}(s)$ cannot vanish on any interval $(s_0-\varepsilon,s_0+\varepsilon)$. To see this we look at the Euler-Lagrange equation obtained by varying $r$,
\begin{equation}
\frac{\rmd}{\rmd s}\frac{\p L}{\p \dot{r}}-\frac{\p L}{\p r} = 0.
\end{equation} 
That is,
\begin{multline}
\frac{\rmd}{\rmd s}\left(\frac{4M}{r}\frac{1}{-1 +\frac{r^2}{l^2}}\dot{t}^* + \frac{2g(r)}{\left(-1 + \frac{r^2}{l^2}\right)^2}\dot{r}\right) - f'(r)\left(\dot{t}^*\right)^2 \\ - \left[\frac{\p}{\p r}\left(\frac{4M}{r}\frac{1}{-1 +\frac{r^2}{l^2}}\dot{t}^* + \frac{g(r)}{\left(-1+\frac{r^2}{l^2}\right)^2}\dot{r}\right)\right]\dot{r} - 2r(\dot{\sigma}^2+\sinh^2(\sigma)\dot{\phi}^2) = 0.
\end{multline}
or, expanding the first term using the chain rule 
\begin{multline}
\left[\frac{\p}{\p r}\left(\frac{4M}{r}\frac{1}{-1 +\frac{r^2}{l^2}}\dot{t}^* + \frac{2g(r)}{\left(-1 + \frac{r^2}{l^2}\right)^2}\right)\right]\dot{r} + \frac{4M}{r}\frac{1}{-1 +\frac{r^2}{l^2}}\ddot{t}^* +\frac{g(r)}{\left(-1+\frac{r^2}{l^2}\right)^2}\ddot{r}- f'(r)\left(\dot{t}^*\right)^2 \\ - \left[\frac{\p}{\p r}\left(\frac{4M}{r}\frac{1}{-1 +\frac{r^2}{l^2}}\dot{t}^* + \frac{g(r)}{\left(-1+\frac{r^2}{l^2}\right)^2}\dot{r}\right)\right]\dot{r} - 2r(\dot{\sigma}^2+\sinh^2(\sigma)\dot{\phi}^2) = 0.
\end{multline}
If $\dot{r}(s)=0$ for all $s \in (s_0-\varepsilon, s_0 +\varepsilon)$, then the two terms in the square brackets don't contribute, and nor does the $\ddot{r}$ term, so we're left with
\begin{equation}
\frac{4M}{r}\frac{1}{-1 +\frac{r^2}{l^2}}\ddot{t}^*- f'(r)\left(\dot{t}^*\right)^2 - 2r(\dot{\sigma}^2+\sinh^2(\sigma)\dot{\phi}^2) = 0,
\end{equation} 
which simplifies to 
\begin{equation}
\frac{4M}{r}\frac{1}{-1 +\frac{r^2}{l^2}}\ddot{t}^*- f'(r)\left(\dot{t}^*\right)^2 -\frac{2h^2}{r^3} = 0.
\end{equation}
Since $E=f(r)\dot{t}^*$ when $\dot{r}=0$, we have that $\dot{t}^*(s) = E/f(r_0)$ for all $s \in (s_0-\varepsilon, s_0 +\varepsilon)$ (where $r_0=r(s_0)$). This is independent of $s$, so $\ddot{t}^*(s)=0$ for all $s \in (s_0-\varepsilon, s_0 +\varepsilon)$. Plugging this in,
\begin{equation}
E^2\frac{f'(r)}{f(r)^2} =\frac{2h^2}{r^3}.
\end{equation}
Now, $rf'(r)=2M/r+ 2r^2/l^2$, and $h^2f(r)/r^2=E^2$ so
\begin{equation}
2E^2\left(\frac{M}{r} + \frac{r^2}{l^2}\right)=2E^2f(r).
\end{equation}
The constant $E$ cannot be $0$, since if it were it must be the case that $\dot{t}^*\equiv 0$, and also that $h=0$. This in turn implies that $\dot{\sigma} \equiv 0$, $\dot{\phi} \equiv 0$, and so $\dot{\gamma}=0$, contradicting the fact that $\dot{\gamma}$ is a null vector field along $\gamma$. Therefore we can divide by $2E^2$ and rearrange to get
\begin{equation}
r(s)=-3M \text{ for all $s \in (s_0-\varepsilon, s_0 +\varepsilon)$ },
\end{equation}
which is absurd, since $r(s)\geq r_+$ by definition. It is therefore impossible for $\dot{r}$ to vanish on any open interval of the parameter $s$, as asserted.
\begin{rmk}
	If we allowed the black hole mass $M$ to be negative, this does not immediately give a contradiction. However, it is quite easy to check that if $M^{\text{ext}}\leq M\leq0$, then $-3M\leq r_+$, and so there are still no constant $r$ null geodesics in the exterior of the black hole.
\end{rmk}
\subsection{Time taken to cross the horizon}
Consider an outgoing null geodesic (that is, one for which $\dot{r} >0$). From the equation of motion it follows that 
\begin{equation}
\begin{split}
\dot{r}&=\sqrt{E^2 - \frac{h^2}{r^2}f(r)},
\\ &=E\sqrt{1-\frac{h^2}{E^2r^2}f(r)}.
\end{split}
\end{equation}
Recalling the definition of $E$, 
\begin{equation}
E = f(r)\dot{t}^* - \frac{2M}{r} \frac{1}{-1+\frac{r^2}{l^2}}\dot{r},
\end{equation} 
we see that
\begin{equation}
\dot{t}^* =\frac{1}{f(r)}\left(E +\frac{2M}{r} \frac{1}{-1+\frac{r^2}{l^2}}\dot{r} \right).
\end{equation}
Dividing $\dot{t}^*$ by $\dot{r}$ we get
\begin{equation}
\begin{split}
\frac{\rmd t^*}{\rmd r} = \frac{\dot{t}^*}{\dot{r}} &= \frac{E}{f(r)\dot{r}} + \frac{2M}{rf(r)}\frac{1}{-1 + \frac{r^2}{l^2}},
\\&=\frac{1}{f(r)}\frac{1}{\sqrt{1-\frac{h^2}{E^2r^2}f(r)}} + \frac{2M}{rf(r)}\frac{1}{-1 + \frac{r^2}{l^2}},
\\&= \frac{1}{f(r)}\left(\frac{1}{\sqrt{1-\frac{\alpha^2}{r^2}f(r)}} + \frac{2M}{r}\frac{1}{-1 + \frac{r^2}{l^2}}\right),
\end{split}
\end{equation}
where in the last line we have defined $\alpha^2=h^2/E^2$. We can now integrate this equation from $r_0$ to $r$ to get $t$ as a function of $r$, provided that the denominator does not vanish for any $r' \in (r_0, r)$, thus
\begin{equation}
\begin{split}
t^*(r)-t^*_0 &= \int_{r_0}^{r}\frac{\rmd t^*}{\rmd r}\, \rmd r,
\\ &= \int_{r_0}^{r}\frac{1}{f(r')}\left(\frac{1}{\sqrt{1-\frac{\alpha^2}{r'^2}f(r')}} + \frac{2M}{r'}\frac{1}{-1 + \frac{r'^2}{l^2}}\right)\, \rmd r'.
\end{split}
\end{equation}
There are three cases, according as $h^2/E^2<l^2$, $h^2/E^2>l^2$, or $h^2/E^2=l^2$. 
\begin{description}
	\item [1. $h^2/E^2<l^2$:]
	In this case, since $f(r)/r^2 < 1/l^2$, we have that 
	\begin{equation}
	1-\frac{\alpha^2}{r^2}f(r) > 1- \frac{h^2}{E^2l^2}>0 \text{ (for all $r>r_+$)}.
	\end{equation}
	Looking at the asymptotic behaviour, we see that 
	\begin{equation}
	\begin{split}
	1-\frac{\alpha^2}{r^2}f(r)&= 1 - \frac{h^2}{E^2l^2}\left(1-\frac{l^2}{r^2}-\frac{2Ml}{r^3}\right),
	\\ & \sim 1-\frac{h^2}{E^2l^2} \text{ as $r \too \infty$.}
	\end{split}
	\end{equation}
	Since $f(r) \sim r^2/l^2$ as $r \too \infty$, it follows that 
	\begin{equation}
	f(r)\sqrt{1-\frac{\alpha^2}{r^2}f(r)} \sim \sqrt{1-\frac{h^2}{E^2l^2}}\frac{r^2}{l^2} \text{ as $r \too \infty$},
	\end{equation}
	so $\left(f(r)\sqrt{1-\frac{\alpha^2}{r^2}f(r)}\right)^{-1}$ is integrable. Similarly, 
	\begin{equation}
	\frac{1}{f(r)}\frac{2M}{r}\frac{1}{-1+\frac{r^2}{l^2}} \sim \frac{2Ml^4}{r^5} \text{ as $r \too \infty$},
	\end{equation}
	which is certainly integrable. Therefore,
	\begin{equation}
	\lim_{r\to \infty} t^*(r) < \infty.
	\end{equation} 
	Informally we say that the geodesic reaches $r=\infty$ in finite coordinate time.
	\item [2. $h^2/E^2>l^2$:] 
	In this case there is a unique value $r^*>r_0>r_+$ such that
	\\ $1-\alpha^2f(r_{\text{max}})/r_{\text{max}}^2 = 0$. It is also true that $\dot{r}$ vanishes when $r=r_{\text{max}}$. Now, as $r$ approaches $r_{\text{max}}$, $1-\alpha^2/r^2f(r)$ behaves like a constant multiple of $(r_{\text{max}}-r)$. But $(r_{\text{max}}-r)^{-1/2}$ is integrable near $r_{\text{max}}$, and so in this case $r$ increases to $r_{\text{max}}$, with $t^*(r_{\text{max}})<\infty$, at which point $\dot{r}$ changes sign, and the geodesic becomes ingoing. 
	\item[3. $h^2/E^2=l^2$:] In this case, 
	\begin{equation}
	1-\frac{\alpha^2}{r^2}f(r) = \frac{\alpha^2}{r^2}\left(1+\frac{2M}{r}\right), 
	\end{equation}
	and so 
	\begin{equation}
	\frac{1}{f(r)} \frac{1}{\sqrt{1-\frac{\alpha^2}{r^2}f(r)}} \sim \frac{l^2}{r^2} \frac{r}{\alpha} \sim \frac{l^2}{\alpha}\frac{1}{r} \text{ as $r \too \infty$}.
	\end{equation}
	But this is not integrable! In other words, 
	\begin{equation}
	\lim_{r\to \infty}t^*(r) = \infty.
	\end{equation}
	Informally we say that the geodesic takes an infinite coordinate time to reach $r=\infty$.
\end{description}
If we look instead at an ingoing null geodesic, that is one for which $\dot{r}<0$ then
\begin{equation}
\dot{r}=-E\sqrt{1-\frac{h^2}{E^2r^2}f(r)}.
\end{equation}
Since it must be the case that initially $h^2f(r_0)/r_0^2<E^2$, and since $h^2f(r)/r^2$ is an increasing function of $r$, it follows that $h^2f(r)/r^2\leq E^2$ for all $r \in \left(r_+, r_0\right)$. Therefore for $r \in \left(r_+, r_0\right)$,
\begin{equation}
t^*(r)-t^*_0 = \int_{r}^{r_0}\frac{1}{f(r')}\left(\frac{1}{\sqrt{1-\frac{\alpha^2}{r'^2}f(r')}} - \frac{2M}{r'}\frac{1}{-1 + \frac{r'^2}{l^2}}\right)\, \rmd r'.
\end{equation}
and so the coordinate time at which the null geodesic $\gamma$ crosses the horizon $r=r_+$, $t^*(r_+)$ is finite. Combining these  two results, we see that the only null geodesics which do not cross the horizon in finite coordinate time are those outgoing null geodesics for which $h^2/E^2=l^2$, which spiral out towards future null infinity, but do not reach it in finite coordinate time.

It is also interesting to note the following proposition. 

\begin{prop}
Define the vector field 
\begin{equation}
\begin{split}
X&=r\left(\frac{\p}{\p r} + \frac{2M}{r}\frac{1}{\left(-1+\frac{r^2}{l^2}\right)}\frac{1}{\left(-1 -\frac{2M}{r}+\frac{r^2}{l^2}\right)}\frac{\p}{\p t^*}\right), 
\\ &=r\left(\frac{\p}{\p r}\right)_{\text{Sch}}.
\end{split}
\end{equation}
If $\gamma$ is a null geodesic, then $g(X, \dot{\gamma})$ is monotonically decreasing along $\gamma$. That is, $\dot{\gamma}\left(g(X, \dot{\gamma})\right)<0$.
\end{prop}
\begin{proof}
Consider,
\begin{equation}
\begin{split}
	\dot{\gamma}\left(g(X, \dot{\gamma})\right) &= \nabla_{\dot{\gamma}}\left(X_\mu \dot{\gamma}^\mu \right)
	\\&=\dot{\gamma}^\mu\nabla_{\dot{\gamma}}X_\mu \text{ (since $\gamma$ is a geodesic)}
	\\&=\dot{\gamma}^\mu \dot{\gamma}^\nu \nabla_\nu X_\mu
	\\&=\dot{\gamma}^\mu \dot{\gamma}^\nu \nabla_{(\mu} X_{\nu)}
	\\&=\dot{\gamma}^\mu \dot{\gamma}^\nu\left( \nabla_{(\mu} X_{\nu)}+\beta g_{\mu\nu} \right) \text{ (for any function $\beta$, since $\gamma$ is null)}
\end{split}
\end{equation}
It now suffices to find a function $\beta$ such that the $\nabla_{(\mu} X_{\nu)}+\beta g_{\mu\nu}$ is negative definite. We define the deformation tensor of $X$ by $\Pi^X_{\mu\nu}=\nabla_{(\mu} X_{\nu)}$. A long but straightforward calculation (or an appeal to one's favourite computer algebra package) gives
\begin{equation}
	\Pi^X = -\frac{(Ml^2+r^3)}{l^2r}\left(\rmd t^*\right)^2 +\frac{l^4r(3M+r)}{p(r)^2}\rmd r^2 + r^2\left(\rmd \sigma^2 +\sinh^2(\sigma)\rmd \phi^2\right).
\end{equation}
If we set $\beta_0(r)=\frac{Ml^2+r^3}{p(r)}$, then the $\left(\rmd t^*\right)^2$ terms cancel, leaving
\begin{equation}
	\Pi^X + \beta_0 g = -\left(\frac{l^2r(r^3+l^2r+4Ml^2)}{p(r)^2}\rmd r^2 + \frac{l^2r^2(3M +r)}{p(r)}\left(\rmd \sigma^2 +\sinh^2(\sigma)\rmd \phi^2\right)\right).
\end{equation}
We can now put $\beta=\beta_0+\delta\beta$, where $\delta\beta$ is positive, so that the coefficient of $\left(\rmd t^*\right)^2$ is negative, but sufficiently small that the other terms remain negative. A suitable choice is $\delta\beta(r)=l^2r/p(r)$, giving
\begin{equation}
	\Pi^X + \beta g =-\left(\rmd t^2 + \frac{l^2r\left(r^3+4Ml^2\right)}{p(r)^2}\rmd r^2+\frac{3Ml^2}{p(r)}r^2\left(\rmd \sigma^2 + \sinh^2(\sigma)\rmd \phi^2\right)\right),
\end{equation} 
which is negative definite. This completes the proof.
\end{proof}
\section{The Klein-Gordon equation}
 The study of the Klein-Gordon in the rest of this paper closely follows the methods and structure of \cite{Boundedness&Growth} and \cite{ToroidalAdS}. Throughout we will assume that $\psi$ is smooth, but this restriction can be lifted by a density argument. 
\subsection{Twisted derivatives}\label{TwistedDerivatives}
 For a given smooth, positive function $f:\mathcal{M}\too \mathbb{R}$, which we shall refer to as a \emph{twisting function}, we define the \emph{twisted derivative} of a function $\psi:\mathcal{M}\too\mathbb{R}$ to be
\begin{equation}
	\td_\mu\psi = f\nabla_\mu\left(\frac{\psi}{f}\right).
\end{equation}
Note that the twisted ``derivative'' is not in fact a derivation; that is, it does not obey the product rule. This manifests itself in the fact that 
\begin{equation}
	\td_\mu\left(1\right) = -\frac{\nabla_\mu f}{f}
\end{equation}
which is not in general zero, but
\begin{equation}
	\td_\mu f =0.
\end{equation}
However the twisted derivative is a tensor, and as with the covariant derivative it satisfies 
\begin{equation}
	\td_\rho\left(g_{\mu\nu}\psi\right)=g_{\mu\nu}\td_\rho\psi.
\end{equation}
If $f$ and $\psi$ have the same asymptotic behaviour as  $r\too \infty$, then the leading order term in $\psi/f$ is annihilated by the derivative, and $\td\psi$ will decay more rapidly than $\nabla\psi$. This will allow the use of energy methods which would otherwise be prevented by the slow decay of some solutions towards infinity. Specifically, the standard energy integrals of solutions satisfying Neumann boundary conditions do not converge, but the twisted energy integrals do. (See for example the introduction of \cite{MassiveWaveEqn}). To this end we will choose 
\begin{equation}
	f(r) \sim r^{-\frac{3}{2}+\kappa}
\end{equation}
which is the asymptotic behaviour of a solution obeying Neumann boundary conditions (see for example \cite{MassiveWaveEqn}, \cite{ToroidalAdS}). 
\paragraph{}The formal $L^2$ adjoint of $\td$, denoted $\td^\dagger$, is given by
\begin{equation}
	\td_\mu^\dagger\psi=-\frac{1}{f}\nabla_\mu\left(f\psi\right).
\end{equation} 
Note that
\begin{equation}
\begin{split}
-\td^{\dagger}_\mu\td^{\mu}\psi &=\frac{1}{f}\nabla_{\mu}\left(f^2\nabla^{\mu}\left(\frac{\psi}{f}\right)\right),
\\ &= \frac{1}{f}\nabla_{\mu}\left(f\nabla^\mu \psi - \psi\nabla^\mu f\right),
\\&= \Box_g \psi - \psi \frac{\Box_g f}{f},
\end{split}
\end{equation}
so we can rewrite the Klein-Gordon equation in terms of twisted derivatives as 
\begin{equation}\label{TwistedKG}
\td^{\dagger}_\mu\td^{\mu}\psi +V\psi =0,
\end{equation} 
where 
\begin{equation}
V = - \left(\frac{\Box_gf}{f}+\frac{\alpha}{l^2}\right).
\end{equation}
We refer to the function $V$ as the \emph{potential} associated with twisting function $f$.

\subsection{The twisted energy-momentum tensor}
For a sufficiently regular function $\psi:\mathcal{M}\too\mathbb{R}$, define the \emph{twisted energy-momentum tensor} 
\begin{equation}
\widetilde{T}_{\mu\nu}[\psi] = \td_\mu\psi\td_\nu\psi - \frac{1}{2}g_{\mu\nu}\left(\td_\sigma\psi\td^\sigma\psi + V\psi^2\right).
\end{equation}
This is the same as the usual definition of the energy momentum tensor of a massive scalar field, except that the derivatives are replaced by twisted derivatives, and the mass term $m^2\psi^2$ is replaced by $V\psi^2$. It is clearly a symmetric tensor of type $(0,2)$ but unlike the usual energy momentum tensor it is not in general divergence free. The following basic properties of the twisted energy momentum tensor were proved in \cite{Boundedness&Growth}. 

\begin{prop}\label{CommonOrGardenProperties}
\begin{enumerate}

\item If $\psi\in\mathcal{C}^2\left(\mathcal{M}\to\mathbb{R}\right)$, then
\begin{equation}
\nabla_\mu\widetilde{T}^\mu{}_{\nu}[\psi] = \left(-\td_\mu^\dagger\td^\mu\psi-V\psi\right)\td_\nu\psi +\widetilde{S}_\nu[\psi], 
\end{equation}
where 
\begin{equation}
	\widetilde{S}_\nu[\psi] = \frac{\td^\dagger_\nu\left(fV\right)}{2f}\psi^2 + \frac{\td^\dagger_\nu f}{2f}\td_\sigma\psi\td^\sigma\psi.
\end{equation}
\item Let $\psi\in\mathcal{C}^2\left(\mathcal{M}\to\mathbb{R}\right)$ be a solution to the Klein-Gordon equation, and let $X$ be a smooth vector field. Define the twisted vector and scalar currents respectively to be 
\begin{equation}
\begin{split}
\tilde{J}^X_\mu[\psi] &= \widetilde{T}_{\mu\nu}[\psi]X^\nu,
\\\widetilde{K}^X[\psi] &= \Pi^X_{\mu\nu}\widetilde{T}^{\mu\nu}[\psi] +X^\nu \widetilde{S}_\nu[\psi],
\end{split}
\end{equation} 
where $\Pi^X_{\mu\nu}=\nabla_{(\mu}X_{\nu)}$ is the deformation tensor of $X$. Then
\begin{equation}
	\nabla^\mu\tilde{J}^X_\mu[\psi]=\widetilde{K}^X[\psi].
\end{equation}
\item If the twisting function $f$ is chosen so that the associated potential $V$ is non-negative, then $\widetilde{T}[\psi]$ obeys the dominant energy condition; that is, if $X$ is a future directed causal vector field, then so is $-\tilde{J}^X[\psi]$.
\end{enumerate}
\end{prop}

\begin{rmk}
\begin{enumerate}[(i)]
\item Part 1. of this proposition is analogous to the usual formula for the divergence of the energy-momentum tensor, except for the additional term $\widetilde{S}[\psi]$ which means that $\widetilde{T}[\psi]$ is not divergence free even when $\psi$ is a solution of the Klein-Gordon equation. 
\item If $X$ is a Killing vector field, then $\Pi^X=0$, and if $Xf=0$, then $X^\nu\widetilde{S}_\nu[\psi]=0$. If both of these are true, then $K^X[\psi]=0$ and so $\tilde{J}^X[\psi]$ is a conserved current.
\item Because of part 3 of proposition \ref{CommonOrGardenProperties} above, we shall choose a twisting function $f$ so that $V\geq0$.
\end{enumerate}
\end{rmk}
\subsection{Boundary Conditions}
Define $\kappa=\sqrt{\frac{9}{4}-\alpha}$. Note that the range $0<\alpha<9/4$ corresponds to $3/2>\kappa>0$, and the conformally coupled case $\alpha=2$ corresponds to $\kappa=1/2$. We say that a function $\psi\in\mathcal{C}^1(\mathcal{M}\to\mathbb{R})$ obeys
\begin{enumerate}
	\item Dirichlet boundary conditions at infinity iff $\kappa>0$ and 
	\begin{equation}
	r^{\frac{3}{2}-\kappa}\psi\too 0\text{, as }r\too\infty,
	\end{equation}
	\item Neumann boundary conditions at infinity iff $0<\kappa<1$ and 
	\begin{equation}
	r^{\frac{5}{2}-\kappa}\td_r\psi\too 0\text{, as }r\too\infty,
	\end{equation}
	\item Robin boundary conditions at infinity iff $0<\kappa<1$ and 
	\begin{equation}
	r^{\frac{5}{2}-\kappa}\td_r\psi+\beta r^{\frac{3}{2}-\kappa}\psi\too 0\text{, as }r\too\infty,
	\end{equation}
	for some non-negative, time-independent, smooth function $\beta$ defined on conformal infinity.
\end{enumerate}
The well-posedness of the Klein-Gordon equation with any of the above types of boundary conditions was established in \cite{MassiveWaveEqn}.
\subsection{An appropriate choice of twisting function}\label{TwistingFunction}
The simplest possible choice of twisting function is $f(r)=r^{-\frac{3}{2}+\kappa}$. After a straightforward calculation, this gives a potential
\begin{equation}
	V(r)=\frac{M\left(3-2\kappa\right)^2}{2r^3}+\frac{3-8\kappa+4\kappa^2}{4r^2}.
\end{equation}
It is easy to check that $3-8\kappa+4\kappa^2\geq0$ for $0<\kappa\leq1/2$, so this is a suitable choice of twisting function for that range (provided that $M\geq0$). Importantly this range includes the confromally coupled case $\kappa=1/2$. As will be explained in section \ref{Instability}, we do not expect to be able to find a twisting function which is suitable for the range $1/2<\kappa<3/2$, for all values of $M>0$.
\begin{rmk}
	The twisted derivative is introduced in order to deal with divergent energy integrals for solutions obeying Neumann and Robin boundary conditions, but is not necessary for Dirichlet boundary conditions (see for example \cite{WellPosednessGH}), so the difficulty in finding an appropriate twisting function for $\kappa>1/2$ is not in principle a barrier to the decay of solutions satisfying Dirichlet conditions.
\end{rmk}

\subsection{Energy boundedness}
\subsubsection{The degenerate energy}
Let 
\begin{equation}
	T=\frac{\p}{\p t^*}.
\end{equation}
Since the metric \eqref{MetricReg} is independent of $t^*$, $T$ is a Killing vector field. Moreover, since we have chosen $f$ to be a function of $r$ only, $Tf=0$. It follows that 
\begin{equation}\label{DivFreeT}
	\nabla^\mu\tilde{J}_\mu^T[\psi]=0
\end{equation} 
whenever $\psi$ is a solution of the Klein-Gordon equation. Let $\mathcal{B}_{[R_1,R_2]}^{[T_1,T_2]}$ denote the region of $\mathcal{M}$ with $T_1\leq t^*\leq T_2$ and $R_1\leq r\leq R_2$, where $0\leq T_1<T_2<\infty$, $r_+<R_1<R_2<\infty$. 
Integrating equation \eqref{DivFreeT} over $B^{[T_1,T_2]}_{[R,_1,R_2]}$ and using the divergence theorem, we get
\begin{equation}\label{DivergenceThm}
\mathcal{\widetilde{E}}_{T_2}[\psi; [R_1,R_2]]-\mathcal{\widetilde{E}}_{T_1}[\psi; [R_1,R_2]]=\mathcal{\widetilde{F}}_{R_2}[\psi; [T_1,T_2]] - \mathcal{\widetilde{F}}_{R_1}[\psi; [T_1,T_2]].
\end{equation}
where
\begin{equation}
\mathcal{\widetilde{E}}_{t^*}[\psi; [R_1,R_2]]=\int_{\Sigma_{t^*}^{[R_1,R_2]}}\tilde{J}^T_\mu[\psi]n^\mu\, \rmd S_{\Sigma_{t^*}}, 
\end{equation}
and
\begin{equation}
\mathcal{\widetilde{F}}_r[\psi; [T_1,T_2]]=\int_{\widetilde{\Sigma}_{r}^{[T_1,T_2]}}\tilde{J}^T_\mu[\psi]m^\mu\, \rmd S_{\widetilde{\Sigma}_{r}}.
\end{equation} 
A direct (but rather long) calculation gives
\begin{equation}\label{T-energy}
\mathcal{\widetilde{E}}_{t^*}[\psi; [R_1,R_2]]=\frac{1}{2}\int_{\Sigma_{t^*}^{[R_1,R_2]}}\left(-g^{t^*t^*}(\p_{t^*} \psi)^2 +g^{rr}(\td_r \psi)^2 + |\slashed {\nabla}\psi|^2 +V\psi^2\right)r^2\,\rmd r\, \rmd \omega,
\end{equation}
and 
\begin{equation}
\mathcal{\widetilde{F}}_r[\psi; [T_1,T_2]]=\int_{\Sigma_{r}^{[T_1,T_2]}}\left(g^{t^*r}(\p_{t^*}\psi)^2+g^{rr}(\p_{t^*}\psi)(\td_r\psi)\right)r^2\,\rmd t^*\, \rmd\omega.
\end{equation}
(Recall that $\slashed\nabla$ is the covariant derivative on the hyperbolic plane.)
\begin{rmk}
	The formula \eqref{T-energy} clearly defines a positive-definite energy if $V>0$.
\end{rmk}
In order to get the fluxes over the whole exterior of the black hole, we will take the limits $R_1\too r_+$ and $R_2\too\infty$. Since $g^{rr}\too 0$ as $r\too r_+$, 
\begin{equation}
\begin{split}
F[\psi;[T_1,T_2]]&:=\lim_{r\to r_+}\mathcal{\widetilde{F}}_r[\psi; [T_1,T_2]],
\\&= \int_{\mathcal{H}_{[T_1,T_2]}}g^{t^*r}(\p_{t^*}\psi)^2r_+^2\,\rmd t\, \rmd\omega.
\end{split}
\end{equation}
where $\mathcal{H}_{[T_1,T_2]}$ is the region of the horizon $r=r_+$ where $T_1\leq t\leq T_2$. 
\begin{rmk}
	Note that $g^{t^*r}$ has the same sign as $M$, and hence so does $F[\psi;[T_1,T_2]]$.
\end{rmk}
\paragraph{} If $\psi$ obeys Dirichlet or Neumann boundary conditions, then by counting powers of $r$ in the metric components it can be seen that 
\begin{equation}
	\lim_{r\to \infty}\mathcal{\widetilde{F}}_r[\psi; [T_1,T_2]]=0.
\end{equation}
If instead $\psi$ obeys Robin boundary conditions, then 
\begin{equation}
\begin{split}
	\lim_{r\to \infty}\mathcal{\widetilde{F}}_r[\psi; [T_1,T_2]]&=-	\lim_{r\to \infty}\frac{1}{2l^2}\int_{\widetilde{\Sigma}_{r}^{[T_1,T_2]}}\beta\p_{t^*}\left(\left(r^{-\frac{3}{2}+\kappa}\psi\right)^2\right)\, \rmd t^* \,\rmd\omega,
	\\&=\lim_{r\to \infty}\left[-\frac{1}{2l^2}\int_{\mathbb{H}^2_{T_2, r}/\Gamma}\beta\left(r^{-\frac{3}{2}+\kappa}\psi\right)^2\,\rmd \omega + \frac{1}{2l^2}\int_{\mathbb{H}^2_{T_1, r}/\Gamma}\beta\left(r^{-\frac{3}{2}+\kappa}\psi\right)^2\,\rmd \omega\right],
	\\ & \qquad\qquad\qquad\qquad\qquad\text{(having carried out the integral in $t^*$)}
	\\&=-\frac{1}{2l^2}\int_{\mathbb{H}^2_{T_2, \infty}/\Gamma}\beta\left(r^{-\frac{3}{2}+\kappa}\psi\right)^2\,\rmd \omega + \frac{1}{2l^2}\int_{\mathbb{H}^2_{T_1, \infty}/\Gamma}\beta\left(r^{-\frac{3}{2}+\kappa}\psi\right)^2\,\rmd \omega.
\end{split}
\end{equation}
Now define the \emph{renormalised energy of $\psi$ at time $t^*$} to be
\begin{equation}\label{RenormalisedEnergy}
	E_{t^*}[\psi] = \mathcal{\widetilde{E}}_{t^*}[\psi; [r_+,\infty]] +\frac{1}{2l^2}\int_{\mathbb{H}^2_{t^*, \infty}/\Gamma}\beta\left(r^{-\frac{3}{2}+\kappa}\psi\right)^2\,\rmd \omega,
\end{equation}
where it is understood that if $\psi$ does not obey Robin boundary conditions, then $\beta \equiv 0$. Since it is assumed that $\beta\geq0$, $E_{t^*}$ is positive definite whenever $\mathcal{\widetilde{E}}_{t^*}$ is.

\paragraph{}Taking the limits $R_1\too r_+$ and $R_2\too\infty$ in \eqref{DivergenceThm}, we get
\begin{equation}
\begin{split}
E_{T_2}[\psi]&=E_{T_1}[\psi] - F[\psi; [T_1,T_2]]. 
\end{split}
\end{equation}
It follows that if $M\geq0$, then
\begin{equation}\label{EnergyDecreasing}
	E_{T_2}[\psi]\leq E_{T_1}[\psi],
\end{equation}
and if $M\leq0$, then
\begin{equation}\label{EnergyIncreasing}
	E_{T_2}[\psi] \geq E_{T_1}[\psi].
\end{equation}
(And if $M=0$, then $E_{T_2}[\psi] = E_{T_1}[\psi]$).

We see that there are two cases to be considered, depending on whether the black hole mass $M$ is positive or negative. When $M>0$, the function $t^* \mapsto E_{t^*}[\psi]$ is non-increasing. In particular it is bounded above by its initial value. It is in this setting that we will prove that for $0<\kappa\leq1/2$, the energy decays polynomially (see Theorem 6.1). On the other hand, when $M\leq0$, the function $t^*\mapsto E_{t^*}[\psi]$ is non-decreasing, and in particular is bounded below by its initial value. 
\begin{rmk}
	In the case $M>0$, if a solution $\psi$ has a negative initial energy then its energy will remain bounded away from zero for all time, and similarly if $M<0$ a solution with \emph{positive} initial energy will remain bounded away from zero. When $M=0$, any solution with non-zero energy will remain bounded away from zero. In section 5, we shall use the existence of such solutions to demonstrate linear instability in certain regions of the parameter space (as illustrated in figure 1) 
\end{rmk}
  
\section{Linear instability}\label{Instability}
Having proved energy boundedness \eqref{EnergyDecreasing}, we are in a position to show that, provided $M\geq0$, the existence of negative energy initial data for the Klein-Gordon equation implies linear instability, and we will explain exactly what is meant by linear instability here. Similarly, when $M<0$ we will show that the existence of positive energy initial data implies linear instability. We will then find a region in the parameter space in which such initial data exist.
\subsection{Negative energy implies instability when $M\geq0$}\label{NegEnergyInst}
We begin by noting that if a solution $\psi$ to the Klein-Gordon equation has negative energy initially
\begin{equation}
E_0[\psi]<0,
\end{equation}
then in view of \eqref{EnergyDecreasing} it has negative energy for all times $t^*\geq 0$
\begin{equation}
E_{t^*}[\psi]\leq E_0[\psi]<0.
\end{equation}
In particular it is impossible to have $E_{t^*}[\psi]\too0$ as $t^*\too\infty$, and so we see that the existence of negative energy solutions provides a barrier to decay. More specifically, consider the following result (which appears in \cite{Quasinormalmodes} as Corollary 1.2).
\begin{thm}
	Suppose that solutions of the Klein-Gordon equation, with boundary conditions fixed, on some asymptotically AdS black hole are bounded in $\underline{H}^1\times\underline{L}^2$. Furthermore suppose that there exists no quasinormal mode on the imaginary axis. Then for any solution $\psi$ of the Klein-Gordon equation with initial data in $D^1(\mathcal{A})\cong \underline{H}^2(\Sigma_{0})$, we have 
	\begin{equation}
	\|\psi\|_{\underline{H}^1(\Sigma_{t^*})} + \|\p_{t^*}\psi\|_{\underline{L}^2(\Sigma_{t^*})} \too 0 \text{ as }t^* \too \infty.
	\end{equation} 
\end{thm}
\begin{rmk}
	Here $\underline{H}^1$ and $\underline{L}^2$ are the twisted Sobolev spaces defined in \cite{MassiveWaveEqn}. They are defined in the same way as the usual Sobolev spaces, but with twisted derivatives replacing ordinary partial derivatives.
\end{rmk}
We also note that Lemma A.1 from \cite{Quasinormalmodes} implies that if $s$ is a quasinormal mode on the imaginary axis, then $s=0$. 
\paragraph{}The contrapositive of this theorem, combined with the fact that a negative energy solution cannot tend to zero tells us that if there exists a negative energy solution, then either
\begin{enumerate}
	\item There exists a quasinormal frequency on the imaginary axis, which in view of the above must be zero, giving a solution which is constant in time, or
	\item No uniform boundedness statement holds. 
\end{enumerate}
It is in this sense that we say that the existence of a negative energy solution implies linear instability.
\begin{rmk}
	Exactly the same argument applied to \eqref{EnergyIncreasing} shows that when $M<0$ the existence of a positive energy solution implies linear instability. 
\end{rmk}
\subsection{The existence initial data leading to instability} In this section, we will find a sufficient condition on $M$ and $\kappa$ for the existence of negative energy initial data when $M\geq0$. We moreover show the existence of positive energy initial data whenever $M\leq0$. If $M$ and $\kappa$ obey one of these condition, then we can conclude from the previous section that there is linear instability. 
\paragraph{}Recall that for solutions obeying Dirichlet or Neumann boundary conditions, the initial energy is given by the formula 
\begin{equation}
E_{0}[\psi]=\frac{1}{2}\int_{\Sigma_{0}}\left(-g^{t^*t^*}(\p_{t^*} \psi)^2 +g^{rr}(\td_r \psi)^2 + |\slashed {\nabla}\psi|^2 +V\psi^2\right)r^2\,\rmd r\, \rmd \omega.
\end{equation}
The first three terms in the integrand are clearly non-negative, but the potential 
\begin{equation}
V(r)=\frac{M\left(3-2\kappa\right)^2}{2r^3}+\frac{3-8\kappa+4\kappa^2}{4r^2}
\end{equation}
may be negative when either $\kappa>1/2$ or $M<0$. Setting $\psi(r)=f(r)$ causes the first three terms to vanish and carrying out the integral in $r$, we get
\begin{equation}
E_{0}[\psi]=\frac{1}{4}\text{Vol}\left(\mathbb{H}^2/\Gamma\right)\left(3-2\kappa\right)r_+^{-3+2\kappa}\left(M+\frac{1-2\kappa}{4\left(1-\kappa\right)}r_+\right). 
\end{equation}
For $\kappa<3/2$, the sign of this quantity is determined by the sign of the term in brackets, 
\begin{equation}\label{InstabilityThreshold}
S:=M+\frac{1-2\kappa}{4\left(1-\kappa\right)}r_.
\end{equation}
We can express $M$ in terms of $r_+$ as
\begin{equation}
M=\frac{r_+}{2}\left(\frac{r_+^2}{l^2}-1\right).
\end{equation}
Plugging this into \eqref{InstabilityThreshold} and rearranging slightly, we see that
\begin{equation}
	S=\frac{r_+}{2}\left(\frac{r_+^2}{l^2}- \left(1-\frac{1}{2}\cdot\frac{1-2\kappa}{1-\kappa}\right)\right).
\end{equation} 
Thus $E_{0}[\psi]<0$ if and only if
\begin{equation}
\frac{r_+^2}{l^2}<1-\frac{1}{2}\cdot\frac{1-2\kappa}{1-\kappa}.
\end{equation}
Since $M\geq0$ corresponds to $r_+/l\geq1$, we see that there is linear instability provided that 
\begin{equation}\label{InstabilityThresholdr_+}
1\leq\frac{r_+^2}{l^2}<1-\frac{1}{2}\cdot\frac{1-2\kappa}{1-\kappa}.
\end{equation} 
(Indeed, whenever this inequality is satisfied we can obtain a negative energy solution by solving the equation with initial data $f(r)$, and conclude that we have linear instability.) 
\paragraph{}Similarly $E_0[\psi]>0$ if and only if
\begin{equation}
\frac{r_+^2}{l^2}>1-\frac{1}{2}\cdot\frac{1-2\kappa}{1-\kappa},
\end{equation} 
and hence there will be instability when 
\begin{equation}
1\geq\frac{r_+^2}{l^2}>1-\frac{1}{2}\cdot\frac{1-2\kappa}{1-\kappa}.
\end{equation}
\paragraph{}In fact, by working slightly harder we can find positive energy initial data whenever $M\leq0$. The idea is that since the only term in the energy that can be negative is the one involving $V\psi^2$, we can proceed as follows: let 
\begin{equation}
	\psi(t^*,r,\sigma,\phi) =
	\begin{cases}
		f(r), & \text{for } r > R, \\
		\Psi(t^*,r,\sigma,\phi), & \text{for } r \leq R.
	\end{cases}
\end{equation}
The integral over $r>R$ will involve only the term $V\psi^2$ and so will be negative. However for $R$ sufficiently large we can make it as small as we please. By making $\left|\Psi\right|$ sufficiently small we can also ensure that the integral of the $V\psi^2$ term over $r\leq R$ is, although negative, as small we please. It now suffices to take $\Psi$ to oscillate sufficiently rapidly in the hyperbolic directions so that the integral of $\left|\slashed\nabla\Psi\right|$ outweighs the two negative terms. Then $E_0[\psi]>0$ and we have linear instability as claimed.    
\paragraph{}This is illustrated in Figure \ref{fig:PrettyPicture}. 
\subsection{The case $M=0$}
Before finishing, a few remarks about the physically interesting $M=0$ (or equivalently $r_+=l$) case are in order.
\subsubsection{Linear scalar hair when $M=0$}
It is a simple calculation to check that if $M=0$ and $\kappa=1/2$ (the conformally coupled case), then setting $\psi(r)=1/r$ we have
\begin{equation}
\Box_g \psi +\frac{\alpha}{l^2}\psi=0.
\end{equation}
That is, there is a non-zero time-independent solution to the Klein-Gordon equation (or linear scalar hair, in the language of the physics community) which obeys Neumann boundary conditions. In particular, this solution does not decay in time, so we have linear instability. This is consistent with \eqref{InstabilityThresholdr_+}. 
\subsubsection{Growing modes when $M=0$} In \cite{InstabilityTopBH} growing mode solutions were found in the case $M=0$ and $1/2<\kappa<1$, but only when Neumann boundary conditions are imposed. This agrees with our argument, as putting $M=0$ in \eqref{InstabilityThreshold}, we see that $E_{0}[\psi]<0$ if and only if $1/2<\kappa<1$. However, no growing modes were found in \cite{InstabilityTopBH} for Dirichlet boundary conditions. The question of whether a decay result can be proved for the $M=0$ case when only Dirichlet boundary conditions are imposed remains open, as does the question of decay in the $M=0$, $0<\kappa<1/2$ case.
\section{Decay rates}
In this section we will only consider the case $M>0$ and $0<\kappa\leq1/2$ so that the twisting function $f(r)=r^{-\frac{3}{2}+\kappa}$ gives a positive potential $V$. 
\subsection{Non-degenerate energy boundedness}
\paragraph{}The renormalised energy is degenerate at the horizon, in the sense that as $r\too r_+$, $g^{rr}\too 0$, so that $E_{t^*}[\psi]$ does not control $\left(\td_r \psi \right)^2$. For $M>0$ degeneracy can be removed using the celebrated redshift effect of Dafermos and Rodnianski \cite{LinWaves}. Define the \emph{non-degenerate renormalised energy density of $\psi$} by
\begin{equation}\label{RenormalisedEnergyDensity}
\mathcal{E}[\psi]=\frac{1}{r}\psi^2+r^4\left(\td_r\psi\right)^2 + \left(\p_{t^*}\psi\right)^2 +r^2 \left|\slashed\nabla\psi\right|^2.
\end{equation}
Note that the coefficients in $\mathcal{E}$ have the same asymptotic behaviour as the coefficients in the degenerate energy, but without the degeneracy at the horizon. We then obtain the following theorem, the proof of which is to be found in section $3.3$ of \cite{LinWaves}.
\begin{thm}(Non-degenerate energy boundedness)
	Consider the Klein-Gordon equation \eqref{KGeqn} on the Lorentzian metric \eqref{Metric}, for fixed $M>0$, $l>0$ and $0<\kappa\leq1/2$. There exists a constant $C>0$ (depending on $M$, $l$ and $\kappa$) such that for any solution $\psi$ of the Klein-Gordon equation, obeying Dirichlet, Neumann, or Robin boundary conditions at infinity, and for any $T_1<T_2$,
	\begin{equation}
	\int_{\Sigma_{T_2}}\mathcal{E}[\psi]\,\rmd r \,\rmd\omega \leq C\int_{\Sigma_{T_1}}\mathcal{E}[\psi]\,\rmd r \,\rmd\omega.
	\end{equation}
\end{thm} 
\subsection{The Morawetz estimate}
In this section, we will prove an integrated decay estimate for solutions of the Klein-Gordon equation. To do so, we will use \emph{energy methods}. These methods were first used by Morawetz for the obstacle problem in Minkowski space in \cite{DecayIBVP} and \cite{NonLinKG}. More recently they were applied in the study of Schwarzschild black holes in \cite{RedShift} and \cite{LinWaves}.
\begin{thm}\label{Morawetz}
Consider the Klein-Gordon equation \eqref{KGeqn} on the Lorentzian metric \eqref{Metric}, for fixed $M>0$, $l>0$ and $0<\kappa\leq1/2$. There exists a constant $C>0$ (depending on $M$, $l$ and $\kappa$) such that for any solution of the Klein-Gordon equation with $0<\kappa\leq1/2$, and any $T_1<T_2$,
\begin{equation}
\int_{\mathcal{B}^{[T_1,T_2]}}\left(\frac{1}{r}\psi^2+\frac{1}{r^2}\left(\p_{t^*}\psi\right)^2+r^4\left(\td_r\psi\right)^2+r^2\left|\slashed\nabla\psi\right|^2\right)\,\rmd t^*\,\rmd r\,\rmd\omega \leq C\int_{\Sigma_{T_1}}\mathcal{E}[\psi]\,\rmd r \,\rmd\omega.
\end{equation} 
\end{thm}
The proof of this theorem requires Lemma 5.1 and Lemma 5.2. We shall present the proof of Theorem 5.1 assuming these, and then give the somewhat technical proofs of the two lemmas. 
\begin{lemma}\label{divlemma}
Define vector fields 
\begin{equation}
	X = rh(r)\frac{\p}{\p r},\, Y=k(r)\frac{\p}{\p t^*},
\end{equation}
and set
\begin{equation}
\begin{split}
	J^{(X,Y,w_1,w_2)}_\mu[\psi]&=\widetilde{T}_{\mu\nu}[\psi]X^\nu+w_1(r)\psi\td_\mu\psi+w_2(r)\psi^2X_\mu -\widetilde{T}_{\mu\nu}[\psi]Y^\nu,
	\\&=:J^{(X,w_1,w_2)}_\mu[\psi]-k(r)\tilde{J}^T_\mu[\psi],
\end{split}
\end{equation}
where $h$, $k$, $w_1$, and $w_2$ are arbitrary smooth functions of $r$, to be determined later. We refer to $J^{(X,Y,w_1,w_2)}$ as a \emph{modified current}.
\paragraph{} Then,
	\begin{equation}\label{DivJ}
	\begin{split}
	-\nabla^\mu J^{(X,Y,w_1,w_2)}_\mu[\psi] =& S_{t^*t^*}\left(\p_{t^*}\psi\right)^2 +S_{rr}\left(\td\psi\right)^2+S_{t^*r}\p_{t^*}\psi\td_r\psi
	\\&+S_{\mathbb{H}^2}\left|\slashed\nabla\psi\right|^2+S_{00}\psi^2+S_{0t^*}\psi\p_{t^*}\psi+S_{0r}\psi\td_r\psi,
	\end{split}
	\end{equation}	
where,
\begin{equation}
\begin{split}
	S_{t^*t^*}=&\left[\frac{l^2\left(\left(1-\kappa\right)r^5-\left(1-2\kappa\right)l^2r^3+\left(5-2\kappa\right)l^2Mr^2-\kappa l^4r-\left(1-2\kappa\right)Ml^4\right)h(r)}{r\left(r^2-l^2\right)^3}\right.
	\\&\left.+\frac{r\left(r^2-l^2\right)\left(r^3-l^2r+2Ml^2\right)h'(r)}{2r\left(r^2-l^2\right)^3} + w_1(r)g_{rr} +2g^{t^*r}k'(r)\right],
	\\S_{rr}=&\left[\frac{2\left(\kappa r^3+\left(1-\kappa\right)l^2r+\left(3-2\kappa\right)Ml^2\right)h(r)-3r\left(r^3-l^2r-2Ml^2\right)h'(r)}{2l^2r}\right.
	\\&-w_1(r)g^{rr}\Big],
	\\S_{t^*r}=&\left[\frac{2Ml^2\left(\left(1-\kappa\right)l^2-\left(2-\kappa\right)r^2\right)h(r)-r\left(r^2-l^2\right)h'(r)}{r\left(r^2-l^2\right)^2}-2w_1(r)g^{t^*r}+2f(r)k'(r)\right],
	\\S_{\mathbb{H}^2}=&-\left[\left(1-\kappa\right)h(r)+\frac{rh'(r)}{2}+w_1(r)\right],
\end{split}
\end{equation}
\begin{equation*}
\begin{split}
S_{00}=&\Bigg[\frac{r}{2}h'(r)V(r)-rw_2(r)h'(r)
\\&-\left[\frac{M\left(3-2\kappa\right)^3}{4r^3}+\frac{\left(1-\kappa\right)\left(3-8\kappa+4\kappa^2\right)}{4r^2}+rw_2'(r)+2\kappa w_2(r)\right]h(r)
\\&\qquad\qquad\qquad\qquad-\left[\frac{3-8\kappa+4\kappa^2}{4r^2}+\frac{M\left(3-2\kappa\right)^2}{2r^3}\right]w_1(r)+rw_2(r)h'(r)\Bigg],
\\S_{0t^*}=&-g^{t^*r}w_1'(r), \text{ and}
\\S_{0r}=&- \left[g^{rr}w_1'(r)+2rh(r)w_2(r)\right].
\end{split}
\end{equation*}
\end{lemma}

\begin{lemma}\label{ineqlemma}
Define vector fields $X$, $Y$ and modified current $J^{(X,Y,w_1,w_2)}$ as in lemma \ref{divlemma}. Then, provided that $h(r)=o(r^2)$, $h(r)w_1(r)=o(r^2)$, and $h(r)w_1(r)=o(r)$, $k(r)$ is bounded, and that $w_2(r)=k_2/r^3$, where $0\leq k_2<r_+V(r_+)/2$,
\begin{equation}
	\int_{\mathcal{B}^{[T_1,T_2]}}-\nabla^\mu J^{(X,Y,w_1,w_2)}_\mu [\psi] \,\rmd\text{Vol} \leq C \int_{\Sigma_{T_1}}\mathcal{E}[\psi]\,\rmd r \,\rmd\omega.
\end{equation} 
for some constant $C$ independent of $T_1$ and $T_2$.
\end{lemma}
\begin{proof}(Of Theorem 5.1)
We begin by noting that if we can pick a modified current satisfying the conditions of lemma \ref{ineqlemma}, which also satisfies
\begin{equation}\label{posdefdiv}
	\frac{1}{r}\psi^2+\frac{1}{r^2}\left(\p_{t^*}\psi\right)^2+r^4\left(\td_r\psi\right)^2+r^2\left|\slashed\nabla\psi\right|^2 \leq Cr^2\left(-\nabla^\mu J^{(X,w_1,w_2)}_\mu[\psi] \right)
\end{equation}
for some positive constant $C$, then the proof is complete. For then integrating this inequality with respect to $\rmd t^*\,\rmd r\,\rmd\omega$ and using lemma \ref{ineqlemma}, the result follows. The difficulty in this proof lies in choosing the functions $w_1$, $w_2$, $h$ and $k$ so that these conditions, particularly \eqref{posdefdiv}, all hold. In fact, this will have to be done in two stages, using two modified currents: one to control all but the $\left|\slashed\nabla\psi\right|^2$ terms on the left hand side of \eqref{posdefdiv}, and another to control the remaining term.     

We begin by simplifying matters and setting $h(r)\equiv1$. Because we do not wish to consider the $\left|\slashed\nabla\psi\right|^2$ term, we will take $w_1$ such that this term vanishes. That is, set $w_1(r)\equiv -\left(1-\kappa\right)$. Plugging these choices into \eqref{DivJ}, we get the following slightly simpler equation,
\begin{equation}
\begin{split}
-\nabla^\mu J&^{(X,Y,w_1,w_2)}_\mu[\psi] =\left[\frac{l^2\left(l^2r^3+3Ml^2r^2-l^4r+Ml^4\right)}{r\left(r^2-l^2\right)^3}+2g^{t^*r}k'(r)\right]\left(\p_{t^*}\psi\right)^2
\\&+\left[\frac{\kappa r^3 +\left(1-\kappa\right)l^2r+\left(3-2\kappa\right)Ml^2}{2l^2r}+\left(1-\kappa\right)g^{rr}\right]\left(\td_r\psi\right)^2
\\& +\frac{\left(3-2\kappa\right)\left(4k_2-M\left(3-2\kappa\right)\right)}{4r^3}\psi^2
\\&+ \left[-\frac{2Ml^2\left(\kappa r^2 +\left(1-\kappa\right)l^2\right)}{r\left(r^2-l^2\right)^2}+2f(r)k'(r)\right]\p_{t^*}\psi\td_r\psi +\frac{2k_2}{r^2}\psi\td_r\psi.
\end{split}
\end{equation}
We choose $k'$ so that the coefficient of $\p_{t^*}\psi\td_r\psi$ vanishes. That is, take
\begin{equation}
	k'(r)=\frac{1}{f(r)}\frac{2Ml^2\left(\kappa r^2 +\left(1-\kappa\right)l^2\right)}{r\left(r^2-l^2\right)^2}>0.
\end{equation}
Note that this gives a positive contribution to the $\left(\p_{t^*}\psi\right)^2$ term, and doesn't affect any other terms. The only remaining cross term is $\psi\td_r\psi$, which we will deal with using Young's inequality with an $\varepsilon$. For each fixed value of $r$, we have
\begin{equation}
\left|\psi\td_r\psi\right| \geq -\left(\varepsilon(r)\psi^2+\frac{1}{4\varepsilon(r)}\left(\td_r\psi\right)^2\right).
\end{equation}
Pick $\varepsilon(r)=\varepsilon_0/r$, for some constant $\varepsilon_0>0$. Then 
\begin{equation}
-\frac{2k_2}{r^2}\psi\td_r\psi \geq -\frac{2k_2\varepsilon_0}{r^3}\psi^2 - \frac{k_2}{2\varepsilon_0r}\left(\td_r\psi\right)^2.
\end{equation}
Our next goal is to pick $\varepsilon_0$ so that these terms can be absorbed by the $\psi^2$ and $\left(\td_r\psi\right)^2$ terms respectively. At this stage we will have to treat the cases $0<\kappa<1/2$ and $\kappa=1/2$ separately. To begin with, let us suppose that $0<\kappa<1/2.$ Looking at the $\psi^2$ term first,
\begin{equation}
	\frac{\left(3-2\kappa\right)\left(2k_2-\frac{1}{2}M\left(3-2\kappa\right)\right)}{2r^3} - \frac{2k_2\varepsilon_0}{r^3}=\frac{\left(3-2\kappa\right)\left(\left(2-\delta\right)k_2-\frac{1}{2}M\left(3-2\kappa\right)\right)}{2r^3},
\end{equation}
where we have defined the positive number 
\begin{equation}
	\delta=\frac{4\varepsilon_0}{\left(3-2\kappa\right)}.
\end{equation}
Recall that $k_2$ must be chosen in the range 
\begin{equation}
	0\leq k_2<r_+V(r_+)/2=\frac{M\left(3-2\kappa\right)^2}{4} +\frac{\left(3-8\kappa+4\kappa^2\right)r_+}{8}.
\end{equation}
Since $\kappa<1/2$,
\begin{equation}
	\frac{M\left(3-2\kappa\right)^2}{4}>\frac{M\left(3-2\kappa\right)}{2},
\end{equation}
and therefore there is an $\eta_0>0$ such that 
\begin{equation}
	\frac{M\left(3-2\kappa\right)}{2} + \eta_0<\frac{M\left(3-2\kappa\right)^2}{4}.
\end{equation}
Set
\begin{equation}
	k_2= \frac{M\left(3-2\kappa\right)}{2} + \eta,
\end{equation}
for some $0\leq \eta\leq\eta_0$, and take $\delta=1$. Then
\begin{equation}
	\frac{\left(3-2\kappa\right)\left(\left(2-\delta\right)k_2-\frac{1}{2}M\left(3-2\kappa\right)\right)}{2r^3}=	\frac{\left(3-2\kappa\right)\eta}{2r^3}>0.
\end{equation}
Moreover, looking at the $\left(\td_r\psi\right)^2$ term, note that 
\begin{equation}
\begin{split}
	\frac{k_2}{2\varepsilon_0r}&= \frac{1}{r}\frac{2}{3-2\kappa}\left(\frac{M\left(3-2\kappa\right)}{2}+\eta\right),
	\\&=\frac{M+\eta'}{r},
\end{split}
\end{equation}
where $\eta'=2\eta/\left(3-2\kappa\right)$. Choosing $\eta'<aM/2$,  (where $a>0$ is small) which is possible since $\eta$ may be as small as we please, we see that
\begin{equation}
\begin{split}
	&\frac{\kappa r^3 +\left(1-\kappa\right)l^2r+\left(3-2\kappa\right)Ml^2}{2l^2r}-\frac{k_2}{2\varepsilon_0r},
	\\&>\frac{\kappa r^3 +\left(1-\kappa\right)l^2r+\left(1-a-2\kappa\right)Ml^2}{2l^2r},
	\\&>0,
\end{split}
\end{equation}
providing we choose $0<a<1-2\kappa$, which is possible since $\kappa<1/2$.
Putting this all together, we have that
\begin{equation}
\begin{split}
-\nabla^\mu J&^{(X,Y,w_1,w_2)}_\mu[\psi] =\left[\frac{l^2\left(l^2r^3+3Ml^2r^2-l^4r+Ml^4\right)}{r\left(r^2-l^2\right)^3}+2g^{t^*r}k'(r)\right]\left(\p_{t^*}\psi\right)^2
\\&+\left[\frac{\kappa r^3 +\left(1-\kappa\right)l^2r+\left(1-a-2\kappa\right)Ml^2}{2l^2r}+\left(1-\kappa\right)g^{rr}\right]\left(\td_r\psi\right)^2
\\& +\frac{\left(3-2\kappa\right)\eta}{2r^3}\psi^2.
\end{split}
\end{equation}
It is then easy to see that
\begin{equation}\label{nontangentialderivatesineq}
-\nabla^\mu J^{(X,Y,w_1,w_2)}_\mu[\psi] \geq c\left(\frac{1}{r^4}\left(\p_{t^*}\psi\right)^2 + r^2\left(\td_r\psi\right)^2 +\frac{1}{r^3}\psi^2\right),
\end{equation}
for some sufficiently small constant $c>0$. 
\paragraph{}Now suppose instead that $\kappa=1/2$. This time, the $\psi^2$ term is
\begin{equation}
	\frac{2k_2-M}{r^3}-\frac{2k_2\varepsilon_0}{r^3}=\frac{2k_2\left(1-\varepsilon_0\right)-M}{r^3},
\end{equation}
and we must choose $k_2$ in the range $0\leq k_2<M$. Writing $k_2=\theta M$, where $0\leq \theta<1$, we see that it if we have $\theta\left(1-\varepsilon_0\right)>1/2$, then the coefficient of $\psi^2$ will be positive. Similarly, the coefficient of $\left(\td_r\psi\right)^2$ is
\begin{equation}
	\frac{r^3+l^2r+4Ml^2}{2l^2r}-\frac{k_2}{2\varepsilon_0r}+\frac{1}{2}g_{rr}=\frac{\varepsilon_0\left(r^3+l^2r\right)+Ml^2\left(4\varepsilon_0-\theta\right)}{2l^2\varepsilon_0r},
\end{equation} 
which is certainly positive provided that $4\varepsilon_0-\theta>0$. If we take (for example) $\varepsilon_0=1/4$ and $\theta=5/6$, then both these inequalities hold, and we can again conclude \eqref{nontangentialderivatesineq}. 
\paragraph{}We'll now look for a current which is able to control the $\left|\slashed\nabla\psi\right|^2$. In order to eliminate as many of the other terms as possible, set $h(r)\equiv0$ and $k(r)\equiv 0$. 
This gives a divergence,
\begin{equation}
\begin{split}
-\nabla^\mu J^{(X,Y,w_1,w_2)}_\mu[\psi]=&w_1(r)g^{rr}\left(\p_{t^*}\psi\right)^2 - w_1(r)g^{rr}\left(\td_r\psi\right)^2-2w_1(r)g^{t^*r}\p_{t^*}\psi\td_r\psi 
\\& -w_1(r)\left|\slashed\nabla\psi\right|^2 -g^{t^*r}w_1'(r)\psi\p_{t^*}\psi - g^{rr}w_1'(r)\psi\td_r\psi
\\& - w_1(r)\left[\frac{3-8\kappa+4\kappa^2}{4r^2}+\frac{M\left(3-2\kappa\right)^2}{2r^3}\right]\psi^2.
\end{split}
\end{equation}
We now use the fact that we control $\psi^2$ as well as the squared $t^*$ and $r$ derivatives (with appropriate weights). In order to control a positive quantity, we need to choose $w_1$ to be negative, and in order to have the right weight in front of the $\left(\p_{t^*}\psi\right)^2$ term, we need to $w_1$ to decay at least as fast as $1/r^2$. Taking $w_1(r)=-1/r^2$, it is then easy to see that
\begin{equation}
\frac{1}{r^2}\left|\slashed\nabla\psi\right|^2\leq -\nabla^\mu J_\mu^{0, \hat{w}_1, \hat{w}_2}[\psi] + C\left(\frac{1}{r^4}\left(\p_{t^*}\psi\right)^2 + r^2\left(\td_r\psi\right)^2 +\frac{1}{r^3}\psi^2\right),
\end{equation} 
for some positive constant $C$. Therefore, 
\begin{equation}
\frac{1}{r}\psi^2+\frac{1}{r^2}\left(\p_{t^*}\psi\right)^2+r^4\left(\td_r\psi\right)^2+\left|\slashed\nabla\psi\right|^2 \leq Cr^2\left(-\nabla^\mu J^{(X,Y,w_1,w_2)}_\mu[\psi] -\nabla^\mu J_\mu^{0, \hat{w}_1, \hat{w}_2}[\psi] \right).
\end{equation}
Upon integrating, we have the required inequality.
\end{proof}

\begin{proof} (Of Lemma \ref{divlemma})
As above set
\begin{equation}
X = rh(r)\frac{\p}{\p r}, Y=k(r)\frac{\p}{\p t^*},
\end{equation}
and 
\begin{equation}
J^{(X,Y,w_1,w_2)}_\mu[\psi]=\widetilde{T}_{\mu\nu}[\psi]X^\nu+w_1(r)\psi\td_\mu\psi+w_2(r)\psi^2X_\mu-\widetilde{T}_{\mu\nu}[\psi]Y^\nu.
\end{equation}
This proof is a (rather long) calculation. We will consider each term separately. The first term is,
\begin{equation}
\begin{split}
	\nabla_\mu\left(\widetilde{T}^\mu{}_\nu[\psi]X^\nu\right)&=\widetilde{K}^X[\psi],
	\\&=\Pi^X_{\mu\nu}\widetilde{T}^{\mu\nu}+\widetilde{S}_\nu X^\nu.
\end{split}
\end{equation} 
Using the formula for $\widetilde{S}$ in terms of $f(r)$ and $V(r)$, we calculate
\begin{equation}
\begin{split}
	\widetilde{S}_\nu X^\nu &=rh(r)\widetilde{S}_r,
	\\& =h(r)\left[\left(3-\kappa\right)V(r)\psi^2-\frac{3-8\kappa+4\kappa^2}{8r^2}\psi^2+\frac{3-2\kappa}{2}\td_\sigma\psi\td^\sigma\psi\right].
\end{split}
\end{equation}
Now we'll look at the term involving the deformation tensor. Define a new tensor $Q=\Pi^X-h(r)g$. It turns out that $Q$ has no $\rmd \sigma$ or $\rmd \phi$ terms, and
\begin{equation}
	g_{\mu\nu}Q^{\mu\nu}=-h(r)+rh'(r). 
\end{equation}
It is then a brief calculation to get
\begin{equation}
\begin{split}
Q_{\mu\nu}\widetilde{T}^{\mu\nu}[\psi]=&\left[Q^{\mu\nu}+\frac{1}{2}\left(h(r)-rh'(r)\right)g^{\mu\nu}\right]\td_\mu\psi\td_\nu\psi 
\\&+\frac{1}{2}\left[h(r)-rh'(r)\right]V(r)\psi^2.
\end{split}
\end{equation}
And using the formula for the twisted energy-momentum tensor, we get
\begin{equation}
	g_{\mu\nu}\widetilde{T}^{\mu\nu}[\psi]= -\td_\sigma\psi\td^{\sigma}\psi - 2V(r)\psi^2.
\end{equation} 
Putting these three equations together, we get 
\begin{equation}
\begin{split}
\nabla_\mu\left(\widetilde{T}^\mu{}_\nu[\psi]X^\nu\right)&=Q_{\mu\nu}\widetilde{T}^{\mu\nu}[\psi]+ h(r)g_{\mu\nu}\widetilde{T}^{\mu\nu}[\psi]+\widetilde{S}_\nu X^\nu,
\\&=\left[Q^{\mu\nu}+\frac{1}{2}\left(h(r)-rh'(r)\right)g^{\mu\nu}+\frac{1-2\kappa}{2}h(r)g^{\mu\nu}\right]\td_\mu\psi\td_\nu\psi
\\&+\left[\left(\frac{\left(3-2\kappa\right)}{2}V(r)-\frac{3-8\kappa+4\kappa^2}{8r^2}\right)h(r)-\frac{r}{2}h'(r)V(r)\right]\psi^2.
\end{split}
\end{equation}
Next, we look at the second term,
\begin{equation}
\nabla_{\mu}\left(w_1(r)\psi\td^\mu\psi\right)=w_1(r)\psi\nabla_\mu\td^\mu\psi+\nabla_\mu\left(w_1(r)\psi \right)\td^\mu\psi.
\end{equation}
By a calculation very similar to that in the proof of Lemma 5.2 in \cite{ToroidalAdS}, we get
\begin{equation}
	\nabla_\mu\td^\mu\psi= \left[\frac{3-8\kappa+4\kappa^2}{r^2}+\frac{M(2-2\kappa)^2}{2r^3}\right]\psi + \frac{3-2\kappa}{2r}\td^r\psi.
\end{equation} 
Therefore,
\begin{equation}\label{divparttwo}
\begin{split}
	\nabla_{\mu}\left(w_1(r)\psi\td^\mu\psi\right)&=w_1'(r)\psi\td^r\psi+w_1(r)\p_\mu\psi\td^\mu\psi \\&\qquad+\left[\frac{3-8\kappa+4\kappa^2}{r^2}+\frac{M(2-2\kappa)^2}{2r^3}\right]w_1(r)\psi^2 + \frac{3-2\kappa}{2r}w_1(r)\psi\td^r\psi,
	\\&=w_1'(r)\psi\td^r\psi+w_1(r)\td_\mu\psi\td^\mu\psi -w_1(r)\psi\td_\mu 1\td^\mu\psi \\&\qquad+\left[\frac{3-8\kappa+4\kappa^2}{4r^2}+\frac{M(2-2\kappa)^2}{2r^3}\right]w_1(r)\psi^2 + \frac{3-2\kappa}{2r}w_1(r)\psi\td^r\psi,
\end{split}
\end{equation}
where we have used the formula $\p_\mu\psi=\td_\mu\psi-\psi\td_\mu1$ to express the partial derivatives in terms of twisted derivatives. Notice that
\begin{equation}
	\td_\mu 1\td^\mu\psi = \td_r1\td^r\psi =\frac{3-2\kappa}{2r}\td^r\psi,
\end{equation} 
and that
\begin{equation}
	\td^r\psi = g^{t^*r}\p_{t^*}\psi + g^{rr}\td_r\psi.
\end{equation}
Putting this in \eqref{divparttwo} we get 
\begin{equation}
\begin{split}
	\nabla_{\mu}\left(w_1(r)\psi\td^r\psi\right)&=w_1(r)g^{t^*t^*}\left(\p_{t^*}\psi\right)^2+w_1(r)g^{rr}\left(\td_r\psi\right)^2
	\\&+2w_1(r)g^{t^*r}\p_{t^*}\psi\td_r\psi +w_1(r)\left|\slashed\nabla\psi\right|^2 + w_1'(r)g^{t^*r}\psi\p_{t^*}\psi 
	\\&+ w_1'(r)g^{rr}\psi\td_r\psi +\left[\frac{3-8\kappa+4\kappa^2}{4r^2}+\frac{M(2-2\kappa)^2}{2r^3}\right]w_1(r)\psi^2.
\end{split}
\end{equation}
The third term is 
\begin{equation}
\begin{split}
	\nabla_\mu\left(w_2(r)\psi^2X^\mu\right)&=w_2'(r)\psi^2X^r+2w_2(r)\psi\p_r\psi X^r +w_2(r)\psi^2 \nabla_\mu X^\mu,
	\\&=\left[rw_2'(r)h(r)+2\kappa w_2(r)h(r)+rw_2(r)h'(r)\right]\psi^2+2rw_2(r)h(r)\psi\td_r\psi,
\end{split}
\end{equation}
where we have used the formula
\begin{equation}
	\nabla_\mu X^\mu = \frac{1}{\sqrt{-g}}\p_\mu\left(\sqrt{-g}X^\mu\right)
\end{equation}
to calculate the divergence. For the fourth term, again using the formula for the divergence of the twisted energy-momentum tensor, we have 
\begin{equation}
\begin{split}
 	\nabla^\mu\left(\widetilde{T}_{\mu\nu}[\psi]Y^\nu\right) &= \Pi^{Y}_{\mu\nu}\widetilde{T}^{\mu\nu}+\widetilde{S}_\nu Y^\nu,
 	\\&=\Pi^{Y}_{\mu\nu}\widetilde{T}^{\mu\nu}, \text{ (the second term vanishes since $\p_{t^*}f=0$),}
 	\\&=2g^{t^*r}k'(r)\left(\p_{t^*}\psi\right)^2 + 2f(r)k'(r)\p_{t^*}\psi\td_r\psi.
\end{split} 
\end{equation} 
Putting these four terms together completes the proof.
\end{proof}
\begin{proof} (Of Lemma \ref{ineqlemma})

Integrating $-\nabla^\mu J^{(X,Y,w_1,w_2)}_\mu[\psi]$ and using the divergence theorem, we get
\begin{equation}
\begin{split}
\int_{\mathcal{M}_{[T_1,T_2]}^{[R_1,R_2]}}-\nabla^\mu J^{(X,Y,w_1,w_2)}_\mu[\psi]\,\rmd \text{Vol} &=\int_{\Sigma_{T_2}^{[R_1,R_2]}}J^{(X,Y,w_1,w_2)}_\mu[\psi] n^\mu\, \rmd S_{\Sigma_{T_2}} -\int_{\Sigma_{T_1}^{[R_1,R_2]}}J^{(X,Y,w_1,w_2)}_\mu[\psi] n^\mu\, \rmd S_{\Sigma_{T_1}}
\\&+\int_{\widetilde{\Sigma}_{R_1}^{[T_1,T_2]}}J^{(X,Y,w_1,w_2)}_\mu[\psi]m^\mu\, \rmd S_{\widetilde{\Sigma}_{R_1}} -\int_{\widetilde{\Sigma}_{R_2}^{[T_1,T_2]}}J^{(X,Y,w_1,w_2)}_\mu[\psi]m^\mu\, \rmd S_{\widetilde{\Sigma}_{R_2}}.
\end{split}
\end{equation}
As before, we will take the limits $R_1\too r_+$ and $R_2\too \infty$. Using the formulas \eqref{formulanup} and \eqref{formulamup} for $n^\mu$ and $m^\mu$, it is straightforward to calculate 
\begin{equation}
	J_\mu^{(X,w_1,w_2)}[\psi]n^\mu=0,
\end{equation}
and
\begin{equation}
\begin{split}
	J_\mu^{(X,w_1,w_2)}[\psi]m^\mu=&-\frac{rh(r)}{2}\frac{g^{t^*t^*}}{\sqrt{g^{rr}}}\left(\p_{t^*}\psi\right)^2 +\frac{rh(r)}{2}\sqrt{g^{rr}}\left(\td_r\psi\right)^2
	\\&-\frac{rh(r)}{2\sqrt{g^{rr}}}\left|\slashed\nabla\psi\right|^2+w_1(r)\frac{g^{t^*r}}{\sqrt{g^{rr}}}\psi\p_{t^*}\psi +w_1(r)\sqrt{g^{rr}}\psi\td_r\psi 
	\\&+\frac{rh(r)}{\sqrt{g^{rr}}}\left(w_2(r)-\frac{1}{2}V(r)\right)\psi^2.
\end{split}
\end{equation}
It is now easy to check by counting powers of $r$, using the asymptotic behaviour imposed by Dirichlet, Neumann or Robin boundary conditions, and the assumptions on $h$, $w_1$ and $w_2$ in the statement of the lemma, that 
\begin{equation}
\lim_{R_2\to \infty}\int_{\widetilde{\Sigma}_{R_2}^{[T_1,T_2]}}J^{(X,w_1,w_2)}_\mu[\psi]m^\mu\, \rmd S_{\widetilde{\Sigma}_{R_2}}=0.
\end{equation}
Now, using the fact that $g^{rr}\too\infty$ as $r\too r_+$ so that the $\left(\td_r\psi\right)^2$ and $\psi\td_r\psi$ terms vanish in the limit, we get
\begin{equation}
\begin{split}
	&\lim_{R_1\to r_+}\int_{\widetilde{\Sigma}_{R_1}^{[T_1,T_2]}}J^{(X,w_1,w_2)}_\mu[\psi]m^\mu\, \rmd S_{\widetilde{\Sigma}_{R_1}}
	\\&= \int_{\mathcal{H}^{[T_1,T_2]}}w_1(r_+)g^{t^*r}\psi\p_{t^*}\psi r_+^2\,\rmd t^*\,\rmd\omega
	\\&+ \int_{\mathcal{H}^{[T_1,T_2]}}\left(\frac{r_+h(r_+)}{\left(-1+\frac{r_+^2}{l^2}\right)}g^{t^*r}r_+^2\left(\p_{t^*}\psi\right)^2-\frac{r_+^3h(r_+)}{2}\left|\slashed\nabla\psi\right|^2+h(r_+)\left(r_+^3w_2(r_+)-\frac{r_+^3}{2}V(r_+)\right)\psi^2\right)\,\rmd t^*\,\rmd\omega,
	\\&\leq\int_{\mathcal{H}^{[T_1,T_2]}}w_1(r_+)g^{t^*r}\psi\p_{t^*}\psi r_+^2\,\rmd t^*\,\rmd\omega + \frac{r_+h(r_+)}{\left(-1+\frac{r_+^2}{l^2}\right)}F[\psi;[T_1,T_2]]
	\\&+ \int_{\mathcal{H}^{[T_1,T_2]}}\left(h(r_+)\left(r_+^3w_2(r_+)-\frac{r_+^3}{2}V(r_+)\right)\psi^2\right)\,\rmd t^*\,\rmd\omega.
\end{split}
\end{equation}
By Young's inequality,
\begin{equation}
	w_1(r_+)\psi\p_{t^*}\psi\leq \varepsilon\psi^2+\frac{w_1(r_+)^2\left(\p_{t^*}\psi\right)^2}{4\varepsilon},
\end{equation}
for ant $\varepsilon>0$. Therefore
\begin{equation}
\begin{split}
 	&\lim_{R_1\to r_+}\int_{\widetilde{\Sigma}_{R_1}^{[T_1,T_2]}}J^{(X,w_1,w_2)}_\mu[\psi]m^\mu\, \rmd S_{\widetilde{\Sigma}_{R_1}}
 	\\&\leq \left(\frac{w_1(r_+)^2}{4\varepsilon}+\frac{r_+h(r_+)}{\left(-1+\frac{r_+^2}{l^2}\right)}\right)F[\psi;[T_1,T_2]] \\&+\int_{\mathcal{H}^{[T_1,T_2]}}\left(h(r_+)\left(r_+^3w_2(r_+)-\frac{r_+^3}{2}V(r_+)\right)+\varepsilon g^{t^*r}\right)\psi^2\,\rmd t^*\,\rmd\omega.
\end{split}
\end{equation}
Setting $w_2(r)=k_2/r^3$, with $0\leq k_2 < r_+^3V(r_+)/2$, we see that the second term can be made negative provided we choose $\varepsilon$ sufficiently small. For such an $\varepsilon$, we have therefore
\begin{equation}
	\int_{\mathcal{B}^{[T_1,T_2]}}-\nabla^\mu J^{(X,w_1,w_2)}_\mu[\psi]\,\rmd \text{Vol}  \leq CF[\psi;[T_1,T_2]]\leq CE_{T_1}[\psi].
\end{equation}
Finally, the terms involving $\tilde{J}^T[\psi]$ are bounded exactly as in the proof of energy boundedness. 
\end{proof}
\subsection{Integrated decay without weight loss}
It is possible to restate the Morawetz estimate Theorem \ref{Morawetz} so that the radial weights are the same on both sides of the inequality. The price that is paid is that the right hand side also includes the energy of $\p_{t^*}\psi$. 
\begin{thm}\label{morawetzderivativeloss}
	Consider the Klein-Gordon equation \eqref{KGeqn} on the Lorentzian metric \eqref{Metric}, for fixed $M>0$, $l>0$ and $0<\kappa\leq1/2$. There is a constant $C>0$ (depending on $M$, $l$ and $\kappa$) such that for any smooth solution $\psi$ to the Klein-Gordon equation, satisfying Dirichlet, Neumann or Robin boundary conditions at infinity,  
	\begin{equation}
		\int_{\mathcal{B}^{[T_1,T_2]}}\mathcal{E}[\psi]\,\rmd t^*\,\rmd r\,\rmd\omega \leq C\int_{\Sigma_{T_1}}\left(\mathcal{E}[\psi]+\mathcal{E}[\p_{t^*}\psi]\right)\,\rmd r \,\rmd\omega.
	\end{equation}
\end{thm}
The proof uses the following Hardy-type inequality, which is Lemma 5.3 in \cite{ToroidalAdS}.
\begin{lemma}
	Let $R>r_+$. Then there is a constant $C>0$ such that 
	\begin{equation}\label{Hardy}
		\int_{r_+}^{\infty}\phi^2\,\rmd r \leq C\left(\int_{r_+}^{\infty}\frac{\phi^2}{r}\,\rmd r+\int_{R}^{\infty}\left(\td_r\phi\right)^2r^2\,\rmd r\right)
	\end{equation}
	for all smooth functions $\phi:\left[r_+,\infty\right)\too\mathbb{R}$ such that $\lim_{r\to\infty}\sqrt{r}\phi=0$. 
\end{lemma} 
\begin{proof}(Of Theorem \ref{morawetzderivativeloss})
	From the Hardy Inequality \eqref{Hardy}, and the original Morawetz estimate Theorem \ref{Morawetz}, it follows that 
	\begin{equation}\label{hardyapplied}
	\int_{\mathcal{B}^{[T_1,T_2]}}\psi^2\,\rmd t^*\,\rmd r\,\rmd\omega \leq C\int_{\Sigma_{T_1}}\mathcal{E}[\psi]\,\rmd r \,\rmd\omega.
	\end{equation}
	Since $\left[\p_{t^*},\Box_g+\frac{\alpha}{l^2}\right]=0$, it follows that $\p_{t^*}\psi$ is also a solution of the Klein-Gordon equation, and therefore we can apply equation \eqref{hardyapplied} to get
	\begin{equation}
	\int_{\mathcal{B}^{[T_1,T_2]}}\left(\p_{t^*}\psi\right)^2\,\rmd t^*\,\rmd r\,\rmd\omega \leq C\int_{\Sigma_{T_1}}\mathcal{E}[\p_{t^*}\psi]\,\rmd r \,\rmd\omega.	
	\end{equation}
	The $\left(\td_r\psi\right)^2$ term appears with the same weight on both sides in \eqref{Morawetz}, so the only remaining term to bound is the $\left|\slashed\nabla\psi\right|^2$ term with the correct weight. To do so, we use a modified current as in the proof of Theorem \ref{Morawetz}. We will choose the current so that $\left|\slashed\nabla\psi\right|^2$ has a positive coefficient in minus the divergence, and all the cross terms vanish. The fact that we control $\psi^2$ and all its other derivatives means that we will then be able to control the other derivatives. Using the formula \eqref{DivJ}, and setting $h(r)\equiv 1$, $w_1(r)\equiv -\left(1-2\kappa\right)$, $w_2(r)\equiv0$ and choosing $k(r)$ so that the $\p_{t^*}\psi\td_r\psi$ cross term vanishes. This choice means that the leading order term in the coefficient of $\psi^2$ appears with a positive sign and so can be discarded. The cross terms all vanish, and the remaining terms all have the same weights as in the bounds above. This completes the proof.   
\end{proof}
\subsection{Quantitative decay}
We have now done enough to establish a quantitative decay estimate for a solution of the Klein-Gordon equation.
\begin{thm}
	 Consider the Klein-Gordon equation \eqref{KGeqn} on the Lorentzian metric \eqref{Metric}, for fixed $M>0$, $l>0$ and $0<\kappa\leq1/2$. There is a constant $C>0$ (depending on $M$, $l$ and $\kappa$) such that if $\psi$ is a solution of the Klein-Gordon equation satisfying Dirichlet, Neumann or Robin boundary conditions, then
	 \begin{equation}
	 	\int_{\Sigma_{t}}\mathcal{E}[\psi]\,\rmd r\,\rmd\omega\leq\frac{C}{\left(1+t\right)^n}\sum_{k=0}^n\int_{\Sigma_{0}}\mathcal{E}[\p_{t^*}^k\psi]\,\rmd r\,\rmd\omega,
	 \end{equation} 
	 for any $n\in\mathbb{N}$.
\end{thm}
The proof of this theorem is exactly as in \cite{ToroidalAdS}.

\bibliographystyle{unsrt}
\bibliography{KGEqnAdSBibliography}

\end{document}